\documentclass[journal]{IEEEtran}

\usepackage[T1]{fontenc}
\ifCLASSINFOpdf
\else
\fi
\IEEEoverridecommandlockouts
\usepackage{multicol}
\usepackage{supertabular}
\usepackage{booktabs}
\usepackage{array}

\makeatletter
\let\mcnewpage\newpage
\newcommand{\changenewpage}{%
	\renewcommand\newpage{%
		\if@firstcolumn
		\hrule width\linewidth height0pt
		\columnbreak
		\else
		\mcnewpage
		\fi
}}
\makeatother
\usepackage{subcaption}
\usepackage{cite}
\usepackage{amsmath,amssymb,amsfonts}
\usepackage{algorithmic}
\usepackage{graphicx}
\usepackage{textcomp}
\usepackage{xcolor}
\usepackage{diagbox}
\def\BibTeX{{\rm B\kern-.05em{\sc i\kern-.025em b}\kern-.08em
		T\kern-.1667em\lower.7ex\hbox{E}\kern-.125emX}}
\usepackage{amsmath, amssymb, bm, cite, epsfig, psfrag}
\usepackage{graphicx}
\usepackage{longtable}
\usepackage[margin=0.7in]{geometry}
\usepackage{dblfloatfix}
\usepackage{array}
\usepackage{lipsum}
\usepackage{gensymb}

\newcolumntype{P}[1]{>{\centering\hspace{0pt}}p{#1}}
\newcolumntype{M}[1]{>{\centering\hspace{0pt}}m{#1}}
\newcolumntype{L}{>{\centering\arraybackslash}m{3cm}}
\usepackage{caption}
\usepackage{epstopdf}	
\usepackage{longtable}
\usepackage{booktabs}
\usepackage{bbm}
\usepackage{multirow}
\usepackage{etoolbox}
\usepackage{pbox}
\usepackage{tabu}	
\usepackage{enumerate}
\usepackage{textcomp}
\usepackage{float}
\usepackage{colortbl}
\usepackage{fancyhdr}
\def\PL{\mathrm{PL}}

\usepackage{soul}
\usepackage{bm}
\newtoggle{conference}
\togglefalse{conference} 
\def\argmax{\mathop{\mathrm{arg\,max}}}

\usepackage{cuted}
\fancypagestyle{firststyle}{
	\fancyhf{}
	\fancyhead[L]{S. Ju, Y. Xing, O. Kanhere, and T. S. Rappaport, “Millimeter Wave and Sub-Terahertz Spatial Statistical Channel Model for an Indoor Office Building,” \textit{IEEE Journal on Selected Areas in Communications, Special Issue on TeraHertz Communications and Networking}, pp. 1–15, Second Quarter 2021.}     

}
\begin{document}

\title{Millimeter Wave and Sub-Terahertz Spatial Statistical Channel Model for an Indoor Office Building}

\author{Shihao~Ju,~\IEEEmembership{Student Member,~IEEE,}
        Yunchou~Xing,~\IEEEmembership{Student Member,~IEEE,}\\
        Ojas~Kanhere,~\IEEEmembership{Student Member,~IEEE,}
        and~Theodore~S.~Rappaport,~\IEEEmembership{Fellow,~IEEE}
\thanks{Manuscript received May 30, 2020; first revised November 22, 2020; second revised February 19, 2021; accepted March 1, 2021. This work was supported in part by NOKIA, in part by the NYU WIRELESS Industrial Affiliates Program, in part by the two National Science Foundation (NSF) Research under Grant 1909206 and Grant 2037845. (Corresponding author: Shihao Ju.)}
\thanks{The authors are with the NYU WIRELESS Research Center, NYU Tandon School of Engineering, New York University, Brooklyn, NY 11201 USA (e-mail: shao@nyu.edu, ychou@nyu.edu, ojask@nyu.edu, tsr@nyu.edu).}}

\maketitle
\thispagestyle{firststyle}
\begin{abstract}
Millimeter-wave (mmWave) and sub-Terahertz (THz) frequencies are expected to play a vital role in 6G wireless systems and beyond due to the vast available bandwidth of many tens of GHz. This paper presents an indoor 3-D spatial statistical channel model for mmWave and sub-THz frequencies based on extensive radio propagation measurements at 28 and 140 GHz conducted in an indoor office environment from 2014 to 2020. Omnidirectional and directional path loss models and channel statistics such as the number of time clusters, cluster delays, and cluster powers were derived from over 15,000 measured power delay profiles. The resulting channel statistics show that the number of time clusters follows a Poisson distribution and the number of subpaths within each cluster follows a composite exponential distribution for both LOS and NLOS environments at 28 and 140 GHz. This paper proposes a unified indoor statistical channel model for mmWave and sub-Terahertz frequencies following the mathematical framework of the previous outdoor NYUSIM channel models. A corresponding indoor channel simulator is developed, which can recreate 3-D omnidirectional, directional, and multiple input multiple output (MIMO) channels for arbitrary mmWave and sub-THz carrier frequency up to 150 GHz, signal bandwidth, and antenna beamwidth. The presented statistical channel model and simulator will guide future air-interface, beamforming, and transceiver designs for 6G and beyond.
\end{abstract}

\IEEEpeerreviewmaketitle

\section{Introduction}
Mobile data traffic is increasing rapidly throughout the world and is predicted to reach 77 exabytes per month by 2022 \cite{Cisco17report}. A large proportion of the data traffic increase comes from emerging indoor wireless applications such as 8K ultra high definition streaming, wireless cognition, and centimeter-level position location, which will be enabled by millimeter-wave (mmWave) and sub-Terahertz (THz) wireless systems due to the vast bandwidths in 6G and beyond \cite{Rap19access,Kanhere21CommMag}. 

Severe outdoor-to-indoor (O2I) penetration loss of up to 60 dB at mmWave frequencies is beneficial for deploying isolated indoor mmWave systems from outdoor co-channel cellular systems \cite{Haneda16vtc}. The 60 GHz band has been well studied in the literature \cite{Smulders92,Xu02jsac,Geng09tvt,Wu17tap} and used in the standards IEEE 802.11ad/ay for wireless local area network (WLAN) \cite{80211ad10,80211ay16}. However, only a few indoor channel measurements and modeling works at other emerging frequencies or across a vast swath of spectra, such as 28, 73, and 142 GHz, have been published \cite{Mac15access, Samman16plos}. Accurate channel models over mmWave and sub-THz frequencies are needed for the design and evaluation of 6G wireless communications and beyond \cite{Rap19access}. 

MmWave and THz (i.e., 30 GHz - 3 THz) have distinct propagation characteristics from sub-6 GHz \cite{Rap13access}. MmWaves do not diffract well and become more sensitive to the dynamic blockage by humans due to the short wavelength \cite{Mac17globecom,Deng16diffraction}. Directional, steerable high gain antennas with beamforming techniques are required to compensate for additional path loss within the first meter of propagation distance as the carrier frequency increases \cite{Xing21b}. Thus, time-variant directional channel models are vital for efficient beam tracking and selection algorithms and proper system design and deployment guidelines.

The remainder of the paper is organized as follows. Section \ref{sec:reivew} provides a brief review of the existing works on channel modeling in indoor environments at mmWave and THz frequencies. Section \ref{sec:desp} describes the 28 and 142 GHz measurement systems used in this work and the indoor office environment, as well as the step-by-step measurement procedure. Section \ref{sec:pathloss} presents the directional and omnidirectional path loss data and resulting models, showing that similar path loss exponents were observed at 28 and 142 GHz in the NLOS environment. Section \ref{sec:method} introduces the 3-D spatial statistical channel impulse response (CIR) model for indoor office scenarios, and Section \ref{sec:stat} provides empirical statistics and distribution fitting of channel parameters derived from the 28 and 140 GHz measurement datasets in both line-of-sight (LOS) and non-line-of-sight (NLOS) environments. Simulated secondary channel statistics (i.e., root mean square (RMS) delay spread (DS) and RMS angular spread (AS)) are generated from the NYUSIM indoor channel simulator and compared with the measured values, which yield good agreements. Finally, concluding marks in Section \ref{sec:conclusion} show that the number of time clusters follows a Poisson distribution and the number of subpaths within each cluster follows a composite exponential distribution for both LOS and NLOS environments at 28 and 140 GHz, but the total number of observed subpaths at 140 GHz is much fewer than the number at 28 GHz.

\section{Existing Works on Indoor Channel Models}\label{sec:reivew}
Numerous indoor channel measurements and studies have been focused on sub-6 GHz and 60 GHz \cite{Saleh87,Rap89tap,Rap91tcomm,Smulders92pimrc,Manabe96jsac,Janssen96tcomm,Sato97tap,Spencer00jsac,Zwick02jsac,Xu02jsac,Foerster03channel,Ghassemzadeh04tcomm,Zwick05tvt,Wu17tap}. Saleh and Valenzuela conducted propagation measurements in an office building using radar-like pulses with 10 ns width at 1.5 GHz, and observed that multipath components (MPCs) arrived in clusters \cite{Saleh87}. A cluster-based statistical channel model was proposed, where the cluster arrival time and the subpath arrival time within each cluster were Poisson distributed, and the expected cluster power and subpath power were modeled as exponentially decaying functions of cluster arrival time and subpath arrival time within each cluster, respectively. This modeling approach has been extensively used in the past few decades. Rappaport conducted propagation measurements at 1.3 GHz in factories and showed that MPCs arrived independently rather than in clusters for factory and open plan building which contain reflecting objects spread throughout the workspace \cite{Rap91tcomm}. Other indoor channel measurements and modeling efforts at mmWave frequencies started from the early 1990s, a majority of which were conducted at 60 GHz \cite{Smulders92pimrc,Manabe96jsac,Janssen96tcomm,Sato97tap,Spencer00jsac,Zwick02jsac,Xu02jsac,Foerster03channel,Ghassemzadeh04tcomm,Zwick05tvt,Wu17tap}.

Standard documents such as IEEE 802.11 ad/ay and 3GPP TR 38.901 presented statistical channel models up to 100 GHz for indoor scenarios such as home, office, shopping mall, and factory \cite{80211ad10,80211ay16,3GPP38901r16}. IEEE 802.11 ad/ay channel models adopted a double-directional CIR model for 60 GHz with dual polarizations based on field measurements and complimentary ray-tracing simulations, which provided detailed temporal and angular channel statistics for conference room, cubical environment, and living room \cite{80211ad10,80211ay16}. 3GPP TR 38.901 proposed a unified geometry-based statistical channel model for indoor and outdoor scenarios for frequencies from 0.5 to 100 GHz, where different scenarios have different values of large-scale parameters (i.e., DS, AS, Rician K factor, and shadow fading) which are required in the channel generation procedure \cite{3GPP38901r16}.

THz communication systems will most likely be deployed in indoor environments to support extremely high data rates of over 100 Gbps \cite{Rap19access}. Considering the particular characteristics of THz signals such as high free space path loss (FSPL), high partition loss, dynamic shadowing loss due to human and vehicle blockages, deep understanding of the THz radio propagation channels is critical for 6G and beyond \cite{Jornet11twc,Ma18APL,Priebe13twc,Han2015twc,He17ttst,Wang20arxiv,Jansen08tap}. First of all, atmospheric or molecular absorption induces non-negligible path loss at THz frequencies, especially at several absorption peaks such as 170 and 325 GHz, which causes about 100 dB/km attenuation \cite{ITU16atmos}. Thus, a frequency-dependent atmospheric absorption term $e^{-k(f)d}$ was introduced in Friis formula \cite{Jornet11twc, Ma18APL}, where $k(f)$ is the atmospheric attenuation factor, $f$ and $d$ are the carrier frequency and transmission distance, respectively. A temporal-spatial stochastic channel model for 275-325 GHz was established based on ray-tracing channel simulations having LOS, first- and second-order reflected paths in an office room \cite{Priebe13twc}. This stochastic channel model can generate channel transfer function, power delay profile (PDP), and angular power spectrum (APS). The adopted ray tracer was calibrated using vector network analyzer (VNA)-based measurements at 275-325 GHz in an office \cite{Priebe13twc}. A generic multi-ray CIR model based on ray tracing consisted of LOS, reflected, diffracted, and scattered paths was proposed and used in channel capacity analysis \cite{Han2015twc}. The reflection, diffraction, and scattering coefficients used in this multi-ray channel model were calibrated by measurements conducted at 0.06-1 THz.  

Most of the existing channel models for THz frequencies were built upon free space, reflection, and scattering measurements for various materials and constructed as a superposition of LOS, reflected and scattered paths in a ray tracing manner. Most propagation measurements were short-distance within a few meters and confined to a single room \cite{Jornet11twc,Ma18APL,Priebe13twc}. 

This paper derives empirical channel statistics based on extensive radio propagation measurements at 28 and 140 GHz conducted on the entire floor of an office building, and proposes a 3GPP-like indoor spatial statistical channel model following the mathematical framework of the NYUSIM outdoor channel models \cite{Samimi16mtt}, which can generate directional and omnidirectional wideband CIRs from 28 to 140 GHz. 

\begin{table}[]
	\centering
	\caption{\textsc{\textcolor{black}{Acronyms}}}
	\begin{tabular}{ l l }
		\hline
		\hline
		\textbf{Acronym} & \textbf{Definition} \\
		\hline
		PDP & Power Delay Profile \\ 
		APS & Angular Power Spectrum  \\  
		FSPL & Free Space Path Loss \\
		CIR & Channel Impulse Response \\
		DS & Delay Spread \\
		AS & Angular Spread \\
		MPC & Multipath Component \\ 
		TC & Time Cluster \\
		SL & Spatial Lobe \\
		MTI & Minimum Void Time Interval \\
		SLT & Spatial Lobe Threshold \\
		HPBW & Half-power Beamwidth \\
		DE & Discrete Exponential Distribution \\
		DU & Discrete Uniform Distribution \\
		RMS & Root Mean Square \\
		\hline
		\hline
	\end{tabular}
	\vspace{-4mm}
\end{table}

\section{28 GHz and 140 GHz Wideband Indoor Channel Measurements} \label{sec:desp}
The 28 and 140 GHz measurement campaigns were conducted in the identical environment, NYU WIRELESS research center on the 9th floor of 2 MetroTech Center in downtown Brooklyn, New York at 2014 and 2019. A wideband sliding correlation-based channel sounder system was used in both measurement campaigns, providing a broad dynamic range of measurable path loss (152 dB at 28 GHz and 145 dB at 140 GHz) \cite{Mac15access,Xing18globecom}. A wideband pseudorandom noise (PN) sequence of length 2047 was generated at baseband, then upconverted to a center frequency of 28 and 142 GHz, and transmitted through a directional and steerable horn antenna at the transmitter (TX). The receiver (RX) captured the RF signal via an identical steerable horn antenna and downconverted and demodulated the RF signal into its baseband I and Q signals \cite{Mac17sounder}. The demodulated signal was then correlated with a local copy of the transmitted signal with a slightly lower rate, which allowed the received signal to ``slide'' past the slower sequence \cite{Mac17sounder}. An average PDP over 20 instantaneous PDPs was sampled by a high-speed oscilloscope and recorded for further analysis. TX and RX antennas were mechanically steered by two electrically-controlled gimbals with sub-degree accuracy in azimuth and elevation planes and were switched between vertical- and horizontal-polarization modes by a 90-degree waveguide twist for co- and cross-polarization studies. The 28 and 140 GHz channel sounder specifications are summarized in Table \ref{tab:sounder_spec}. Null-to-null RF bandwidths of 800 MHz and 1 GHz were adopted in the 28 and 140 GHz measurement campaigns, resulting in a time resolution of MPC equal to 2.5 ns and 2 ns, respectively \cite{Mac15access,Xing18globecom}. 

Omnidirectional channel statistics are often preferred in channel models and channel simulations since arbitrary antenna patterns can be added \cite{Samimi16mtt}.
Thus, omnidirectional PDPs should be recovered from measured directional PDPs by aligning these measured PDPs with absolute time delays \cite{Sun15synthesize,Samimi16mtt,Ju20gc}. However, the 28 GHz channel sounder did not have precise synchronization between TX and RX and cannot provide absolute timing information of measured PDPs since the PDP recording was triggered at the first MPC arrival and only had excess time delay information. A 3-D ray tracer NYURay was employed to provide the time of flight (i.e., absolute time delay) of the first arriving MPC in a measured PDP, which will be explained in Section \ref{sec:synthesis}. The 140 GHz channel sounder was equipped with rubidium standard references at both TX and RX sides for frequency/timing synchronization \cite{Mac17sounder}; however, we used the absolute time delays obtained from the 3-D ray tracer to synthesize omnidirectional PDPs for both 28 and 140 GHz data for processing consistency.

\begin{table}[]
	\caption{\textsc{Specifications for the 28 GHz and 142 GHz sliding correlator channel sounding systems \cite{Mac15access,Xing19globecom}.}}
	\label{tab:sounder_spec}
	\begin{tabular}{|c|c|c|}
		\hline
		\textbf{Carrier Frequency}           & \textbf{28 GHz}            & \textbf{142 GHz}        \\ \hline
		\textbf{Probing Signal}              & \multicolumn{2}{c|}{11th order PRBS (length=2047)}   \\ \hline
		\textbf{TX PN Code Chip Rate}        & 400 Mcps                   & 500 Mcps                \\ \hline
		\textbf{TX PN Code Chip Width}       & 2.5 ns                     & 2.0 ns                    \\ \hline
		\textbf{RX PN Code Chip Rate}        & 399.95 Mcps                & 499.9375 Mcps           \\ \hline
		\textbf{Sliding Factor}              & \multicolumn{2}{c|}{8000}                            \\ \hline
		\textbf{Digitizer Sample Rate}       & 2 Msps               & 2.5 Msps          \\ \hline
		\textbf{RF Bandwidth (Null-to-Null)} & 800 MHz                    & 1 GHz                   \\ \hline
		\textbf{TX/RX Antenna Gain}          & 15 dBi                     & 27 dBi                  \\ \hline
		\textbf{TX/RX Azimuth HPBW}          & 28.8\degree & 8\degree \\ \hline
		\textbf{TX/RX Elevation HPBW}        & 30\degree   & 8\degree \\ \hline
		\textbf{Multipath Resolution}        & 2.5 ns                     & 2 ns                    \\ \hline
		\textbf{Max. Transmit Power}   & 23.9 dBm                     & 0 dBm                  \\ \hline
		\textbf{Max. Measurable Path Loss}   & 152 dB                     & 145 dB                  \\ \hline
		\textbf{TX Polarization}             & \multicolumn{2}{c|}{Vertical}                        \\ \hline
		\textbf{RX Polarization}             & \multicolumn{2}{c|}{Vertical/Horizontal}             \\ \hline
	\end{tabular}
	\vspace{-4mm}
\end{table}

\begin{figure*}[h!]
	\centering
	\includegraphics[width=.9\textwidth]{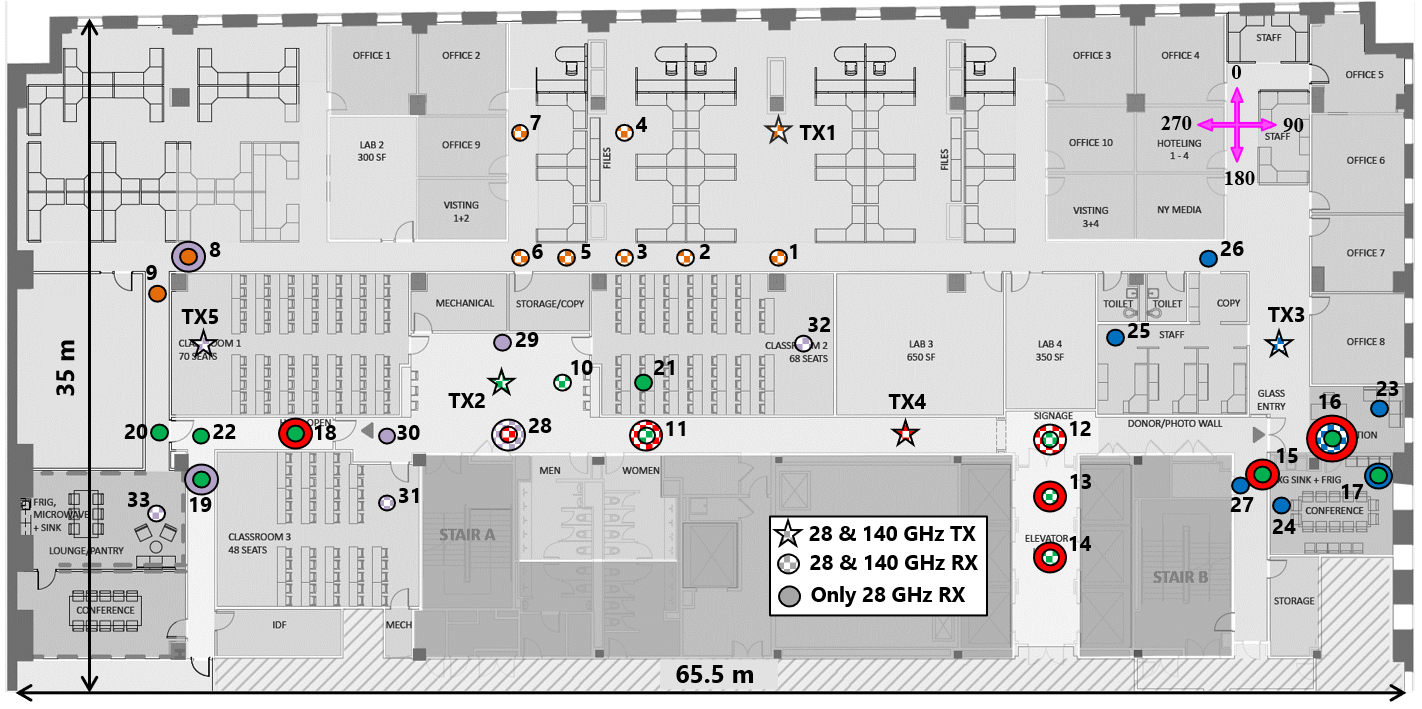}
	\caption{\textcolor{black}{Floor plan of the 9th floor, 2 MetroTech Center \cite{Mac15access}. TX and RX locations measured at both 28 GHz and 140 GHz are denoted as stars and circles with checkerboard texture, respectively. RX locations only measured at 28 GHz are denoted as solid circles. Each of the five TX locations is denoted in a different color, and the RX locations paired with a TX location is denoted in the same color. }}
	\label{fig:floor_plan}
\end{figure*}
\subsection{Measurement Environment and Procedure}\label{sec:measenv}
The measurements were conducted in a typical indoor office environment (65.5 m $\times$ 35 m $\times$ 2.7 m) with offices, conference rooms, classrooms, long hallways, open-plan cubicles, and elevators, as shown in Fig. \ref{fig:floor_plan}. Common obstructions are desks, chairs, cubicle partitions, glass doors, and walls made of drywall with metal studs. 

\textcolor{black}{Five TX locations and 33 RX locations were selected in the 28 GHz measurement campaign in 2014. Overall, measurements were conducted at nine LOS location pairs (i.e., a pair of TX and RX locations) and 35 NLOS location pairs, where the 3-D TX-RX (T-R) separation distances ranged from 3.9 m to 45.9 m. The identical five TX locations and a subset of the RX locations were measured in the 140 GHz measurement campaign due to the limit of maximum transmit power in 2019 and 2020, resulting in nine LOS location pairs and 13 NLOS location pairs. The T-R separation distance ranged from 3.9 m to 39.2 m. In Fig. \ref{fig:floor_plan}, TX and RX locations measured at both 28 GHz and 140 GHz are denoted as stars and circles with checkerboard texture, respectively. RX locations only measured at 28 GHz are denoted as solid circles. Each RX location was paired with the TX location in the same color. One RX location with multiple colors indicates that this RX location was measured with multiple TX locations (e.g., RX16 was measured with TX2, TX3, and TX4).} 


\textcolor{black}{For each T-R location pair, eight unique antenna azimuth sweeps were measured to investigate the spatial statistics of arrival and departure, where six RX antenna azimuth sweeps and two TX antenna azimuth sweeps were performed. During each sweep, TX (RX) horn antenna was rotated in step increments of the antenna half-power beamwidth (HPBW) \cite{Mac15access} so that the directional measurements can emulate channel measurements using omnidirectional antennas. The detailed description of each measurement sweep is listed in Table \ref{tab:sweep}.}

\begin{table*}[]
	\centering
	\caption{\textsc{\textcolor{black}{TX/RX Antenna Sweep Description}}}
	\label{tab:sweep}
	\begin{tabular}{|p{0.05\linewidth} | p{0.08\linewidth}|p{0.75\linewidth}|}
		\hline
		\textbf{Sweep Index} & \textbf{Sweep Type} & \textbf{Description}                                                                                                                                                                                                                                                                \\ \hline
		1                                                                     & RX sweep                                                          & The TX and RX antennas were pointed directly towards each other on boresight in both the azimuth and elevation planes (for LOS or NLOS environments). The RX antenna was then swept in the azimuth plane in steps of HPBW, for a fixed TX antenna at the boresight azimuth and elevation angles.                                                    \\ \hline
		2                                                                     & RX sweep                                                          & With respect to the boresight angle in elevation, the RX antenna was uptilted by HPBW and then swept in the azimuth plane in steps of HPBW, for a fixed TX antenna at the boresight azimuth and elevation angles.                                                                                                                                              \\ \hline
		3                                                                     & RX sweep                                                          & With respect to the boresight angle in elevation, the RX antenna was downtilted by HPBW and then swept in the azimuth plane in steps of HPBW, for a fixed TX antenna at the boresight azimuth and elevation angles.                                                                                                                                         \\ \hline
		4                                                                     & RX sweep                                                          & With respect to the boresight angle in elevation, the TX antenna was uptilted by HPBW. The RX antenna was fixed at the boresight elevation angle, and then swept in the azimuth plane in steps of HPBW.                                                                                                                                                           \\ \hline
		5                                                                     & RX sweep                                                          & With respect to the boresight angle in elevation, the TX antenna was downtilted by HPBW. The RX antenna was fixed at the boresight elevation angle, and then swept in the azimuth plane in steps of HPBW.                                                                                                                                                       \\ \hline
		6                                                                     & TX sweep                                                          & The TX and RX antennas were pointed directly towards each other on boresight in both the azimuth and elevation planes. The TX antenna was then swept in the azimuth plane in steps of HPBW, for a fixed RX antenna at the boresight azimuth and elevation angles.                                                                                 \\ \hline
		7                                                                     & RX sweep                                                          & This measurement was an RX sweep with the TX antenna set to the second strongest AOD in the azimuth and elevation plane. The second strongest AOD was determined by comparing the signal level from all the AODs during Measurement 6, except for the angles corresponding to the main angle of arrival. The RX antenna was fixed at the boresight elevation angle and then swept in steps of HPBW in the azimuth plane. \\ \hline
		8                                                                     & TX sweep                                                          & This measurement corresponds to the second TX sweep with TX antenna either uptilted or downtilted by HPBW after determining the elevation plane with the strongest received power from Measurement 4 and Measurement 5 during measurements. The RX antenna was pointed towards the initial boresight azimuth and elevation angles, and the TX was uptilted or downtilted by HPBW, and then swept in steps of HPBW in the azimuth plane.                                                                                                                                                                                                                                                                                                                                                   \\ \hline
	\end{tabular}
	\vspace{-4mm}
\end{table*}

The equivalent omnidirectional received power can be synthesized by summing the received powers from all measured unique pointing angles obtained at antenna HPBW step increments in both planes \cite{Sun15synthesize}. The sweeping step was equal to the antenna HPBW (30\degree~for 28 GHz and 8\degree~for 140 GHz), which corresponded to 12 and 45 rotation steps over the complete azimuth plane, respectively. At each rotation step, an averaged PDP over 20 instantaneous PDPs with accurate excess timing information with time resolutions of 2.5 and 2 ns was recorded for 28 GHz and 140 GHz, respectively. Note that two antenna polarization configurations, vertical-to-vertical (V-V) and vertical-to-horizontal (V-H), were measured using the identical procedure described above, resulting in 16 measurement sweeps in total at each unique T-R location pair. This paper mainly focuses on the co-polarized (V-V) polarization to develop the omnidirectional and directional indoor channel models. For each T-R location pair, at most 96 (= $8 \times 12$) at 28 GHz and 360 (= $8 \times 45$) at 140 GHz directional PDPs were acquired with V-V polarization configuration.   

\subsection{Synthesizing Omnidirectional PDPs} \label{sec:synthesis}
A 3-D mmWave ray-tracing software, NYURay \cite{Kanhere19globecom}, was used to predict possible propagating rays between the TX and RX and provided the time of flight (i.e., absolute time delay) of the first arriving MPC of a measured directional PDP. Since the horn antennas had beamwidths of 30\degree~and 8\degree~for 28 GHz and 140 GHz, the exact angle of departure and angle of arrival of MPCs were unknown. Each measured directional PDP was assigned to a predicted ray which was closest to this directional PDP in space, then the absolute time delay of the first arriving MPC of the PDP was set to be the time of flight of the corresponding predicted ray. Directional PDPs were aligned in the temporal domain and summed to generate an omnidirectional PDP. Being closest in space between a measured PDP and the set of predicted rays means the antenna gain in the direction of a predicted ray when the antenna is pointing to the direction of the measured PDP is the highest among all predicted rays:
	\begin{equation}
		\argmax_{p\in \textup{P}}  G_{\phi}(\Delta\phi_{\textup{AOD}}) + G_{\theta}(\Delta\theta_{\textup{ZOD}}) + G_{\phi}(\Delta\phi_{\textup{AOA}}) + G_{\theta}(\Delta\theta_{\textup{ZOA}}),
	\end{equation}
where $\textup{P}$ is the set of predicted rays from NYURay. $\Delta\phi_{\textup{AOD}}$, $\Delta\theta_{\textup{ZOD}}$, $\Delta\phi_{\textup{AOA}}$, \textcolor{black}{$\Delta\theta_{\textup{ZOA}}$} denote the absolute difference of the azimuth angle of departure (AOD), zenith angle of departure (ZOD), azimuth angle of arrival (AOA), zenith angle of arrival (ZOA) between the measured directional PDP and a predicted ray, respectively. $G_{\phi}$ and $G_{\theta}$ represent the antenna pattern in the azimuth and elevation planes, respectively. The gain at the peak of the antenna main lobe is normalized to 0 dB. Thus, the antenna gain is -3 dB when the angle difference is 1/2 HPBW from the peak of the main lobe.

MPCs recorded in different directional PDPs may have originated from the same predicted ray, in which case the measured MPC was an antenna-gain weighted version of the true MPC. Thus, the directional PDPs assigned to the same predicted ray were summed in powers and generate a partial-omnidirectional PDP for MPC extraction to avoid double counting. The direction of the extracted MPC was assumed to be the direction of the measured directional PDP which was closest to the predicted ray in space. An omnidirectional PDP was recovered, and MPCs were extracted by applying this procedure to all directional PDPs measured at each T-R location pair. Due to the mismatching between the measured PDPs and predicted rays at a few locations, we recovered omnidirectional PDPs for 37 of 44 location pairs at 28 GHz and 20 of 22 location pairs at 140 GHz. Table \ref{tab:28stat_new} and Table \ref{tab:140stat_new} in the Appendix give TX-RX location pairs used in this paper.


\section{Large-scale Path Loss Models} \label{sec:pathloss}
\subsection{Directional Path Loss Modeling}
Antenna arrays with many elements will enable steerable and narrow beams to compensate for the large free space path loss in the first meter at mmWave and sub-THz frequencies. Directional path loss modeling is increasingly critical for future 6G and beyond communication system design. Thus, in this work, rotatable high-gain horn antennas were used at both the TX and RX during the 28 GHz and 140 GHz measurements, as shown in Table \ref{tab:sounder_spec}, to study double-directional channels. 

We use the close-in free space reference distance (CI) path loss model with 1 m reference distance \cite{Rap13access}, as this has been proven to be superior for modeling path loss over many environments and frequencies \cite{Sun16tvt}. $\PL^\textup{CI}$ represents the path loss in dB scale, which is a function of distance and frequency:
\begin{equation}\label{CI1}
	\begin{split}
		\PL^\textup{CI}(f, d)[\textup{dB}]=&\textup{FSPL}(f, d_0)+10n\log_{10}\left(\frac{d}{d_0}\right)+\chi_{\sigma} \text{,}\\
		&\text{for}\: d\geq d_0, \;\;\text{where}\: d_0 = 1 \textup{m}
	\end{split}
\end{equation}
where $n$ denotes the path loss exponent (PLE), and $\chi_{\sigma}$ is the shadow fading (SF) that is commonly modeled as a lognormal random variable with zero mean and $\sigma$ standard deviation in dB. $d$ is the 3-D T-R separation distance. $d_0$ is the reference distance, and $\textup{FSPL}(f, d_0)=20\log_{10}(4\pi d_0 c/f)$. The CI path loss model uses the FSPL at $d_0 = 1$ m as an anchor point and fits the measured path loss data with a straight line controlled by a single parameter $n$ (PLE) obtained via the minimum mean square error (MMSE) method. 

Throughout this paper, LOS and NLOS locations are defined according to whether the TX and RX can see each other. Here, for the directional path loss modeling, we define the \textit{LOS direction} for LOS locations as the direction when the TX and RX directional antennas are pointed directly to each other. The LOS direction can be calculated based on the relative position of the TX and RX - the LOS direction is along the line of bearing between the TX and RX. The \textit{NLOS-Best direction} is only defined for NLOS locations, and represents the best pointing direction for which minimum path loss is measured, which can be found by thoroughly rotating TX and RX directional antennas in the 3D space. The \textit{NLOS direction} is defined for both LOS and NLOS locations, and represent all pointing directions which received detectable powers other than the LOS and NLOS-Best directions \cite{Mac15access,Deng15icc}.

\begin{figure}
	\centering
	\begin{subfigure}[b]{.5\textwidth}
		\centering
		\includegraphics[width=0.9\linewidth]{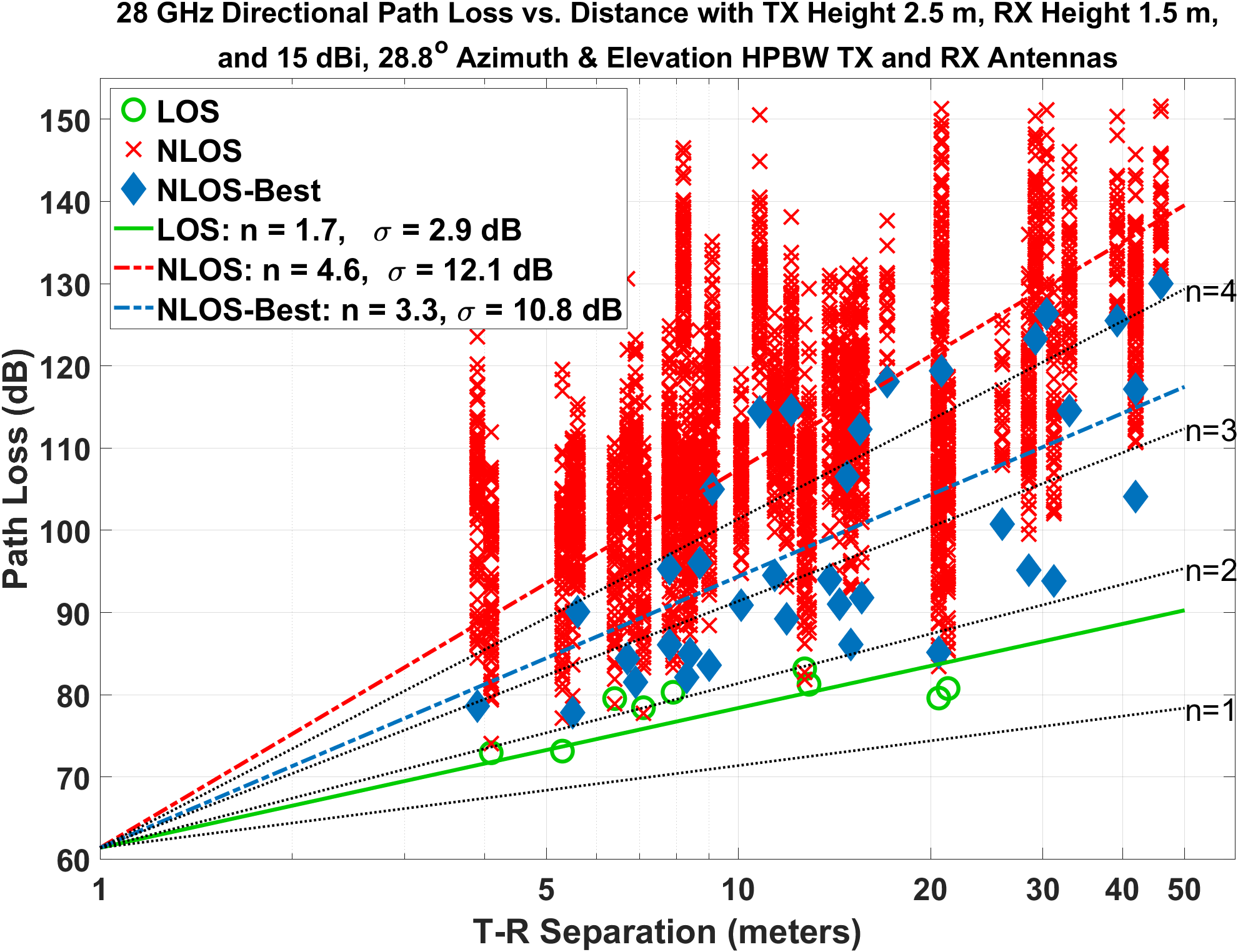}
		\caption{28 GHz indoor directional CI path loss model and data \cite{Mac15access}.}
		\label{fig:dir_pl_28}
	\end{subfigure}
	\hfill
	\vspace{.1em}
	\begin{subfigure}[b]{.5\textwidth}
		\centering
		\includegraphics[width=0.9\linewidth]{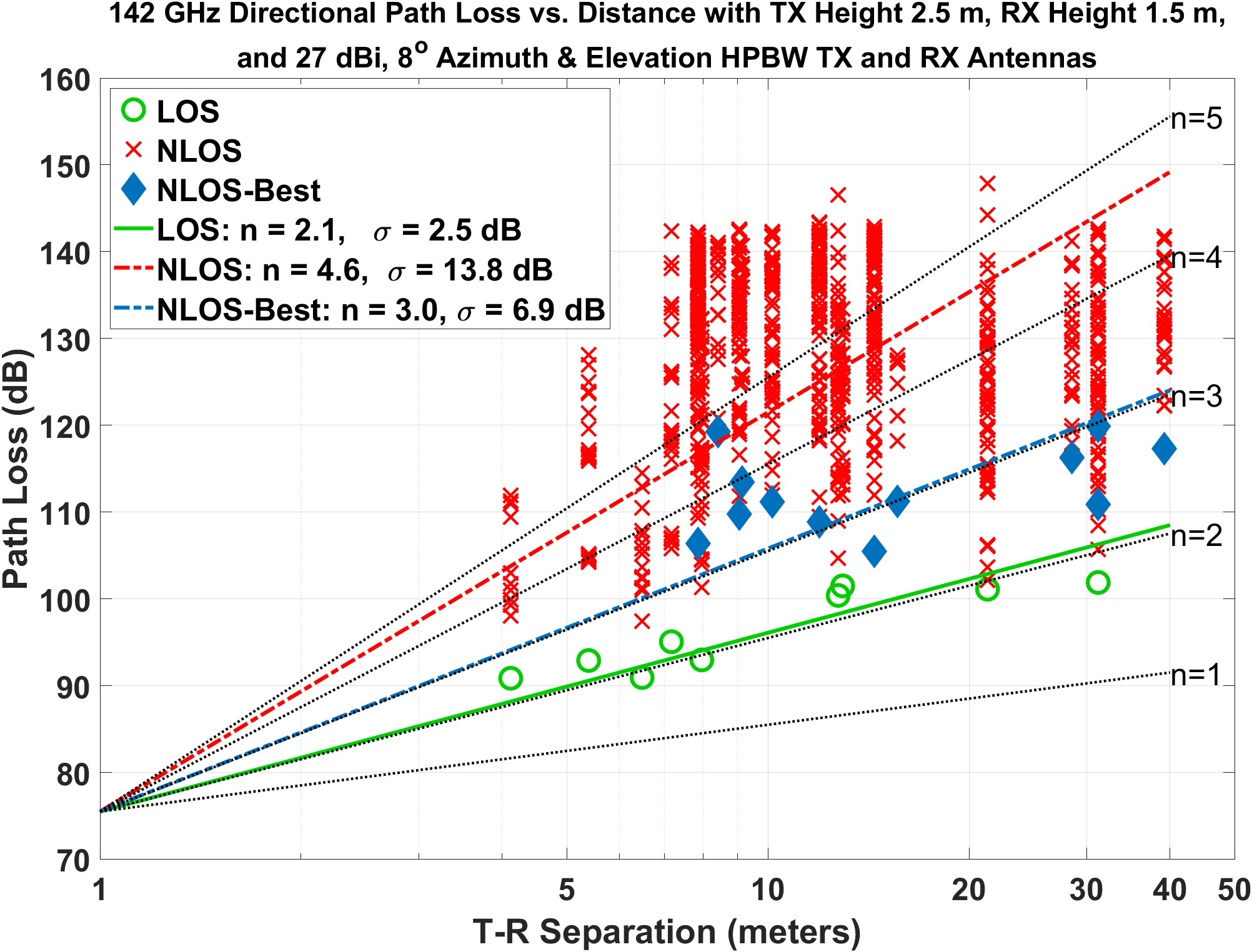}
		\caption{142 GHz indoor directional CI path loss model and data.}
		\label{fig:dir_pl_140}
	\end{subfigure}
	\caption{28 GHz and 142 GHz indoor directional CI path loss models and scatter plots with TX antenna height of 2.5 m and RX antenna height of 1.5 m for V-V polarization.}
	\label{fig:dir_pl}
\end{figure}

Fig. \ref{fig:dir_pl} shows the directional CI path loss model using measured path loss data at 28 GHz and 142 GHz \cite{Mac15access}. The path loss in the LOS direction is represented by a green circle for the LOS locations, and the path loss in the NLOS-Best direction is represented by a blue diamond for the NLOS locations. Measurements pointed to other directions are denoted by red crosses as NLOS directions for both LOS and NLOS locations. Comparing Fig. \ref{fig:dir_pl_28} and Fig. \ref{fig:dir_pl_140}, the LOS PLEs at 28 and 142 GHz are 1.7 and 2.1 which may be due to the differences in antenna HPBWs (30\degree~and 8\degree). Wider beamwidths may capture more energy through reflection and scattering in the vicinity of the LOS direction, causing a PLE of less than 2. Furthermore, the NLOS-Best PLE at 28 and 142 GHz are about 3.0, suggesting strong NLOS paths are available to provide a sufficient link margin and can be leveraged by intelligent reflecting surfaces \cite{Basar19access}.

\subsection{Omnidirectional Path Loss Modeling}
\begin{figure}
	\centering
	\begin{subfigure}[b]{.5\textwidth}
		\centering
		\includegraphics[width=0.9\linewidth]{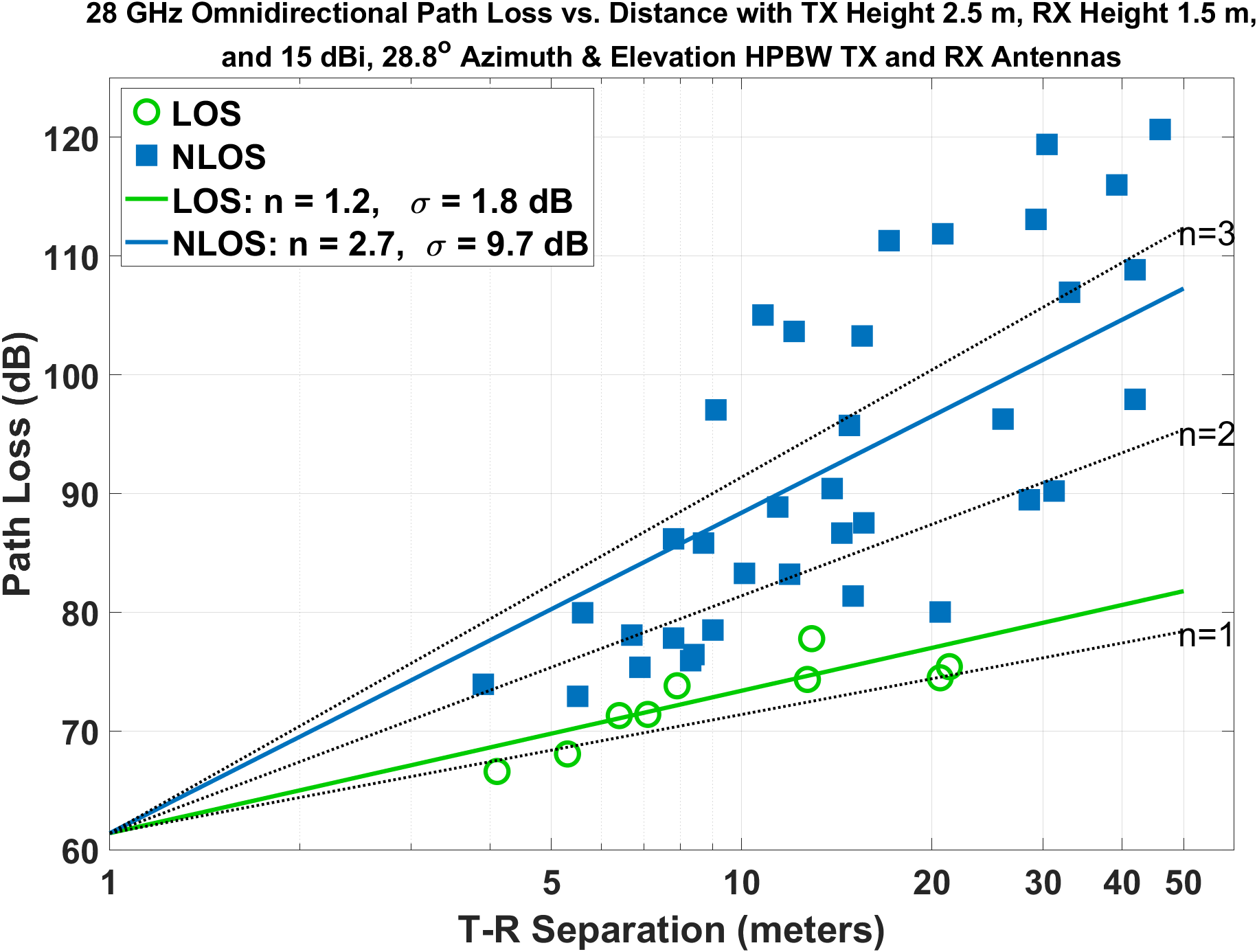}
		\caption{28 GHz indoor omnidirectional CI path loss model and data \cite{Mac15access}.}
		\label{fig:omni_pl_28}
	\end{subfigure}
	\hfill
	\vspace{.1em}
	\begin{subfigure}[b]{.5\textwidth}
		\centering
		\includegraphics[width=0.9\linewidth]{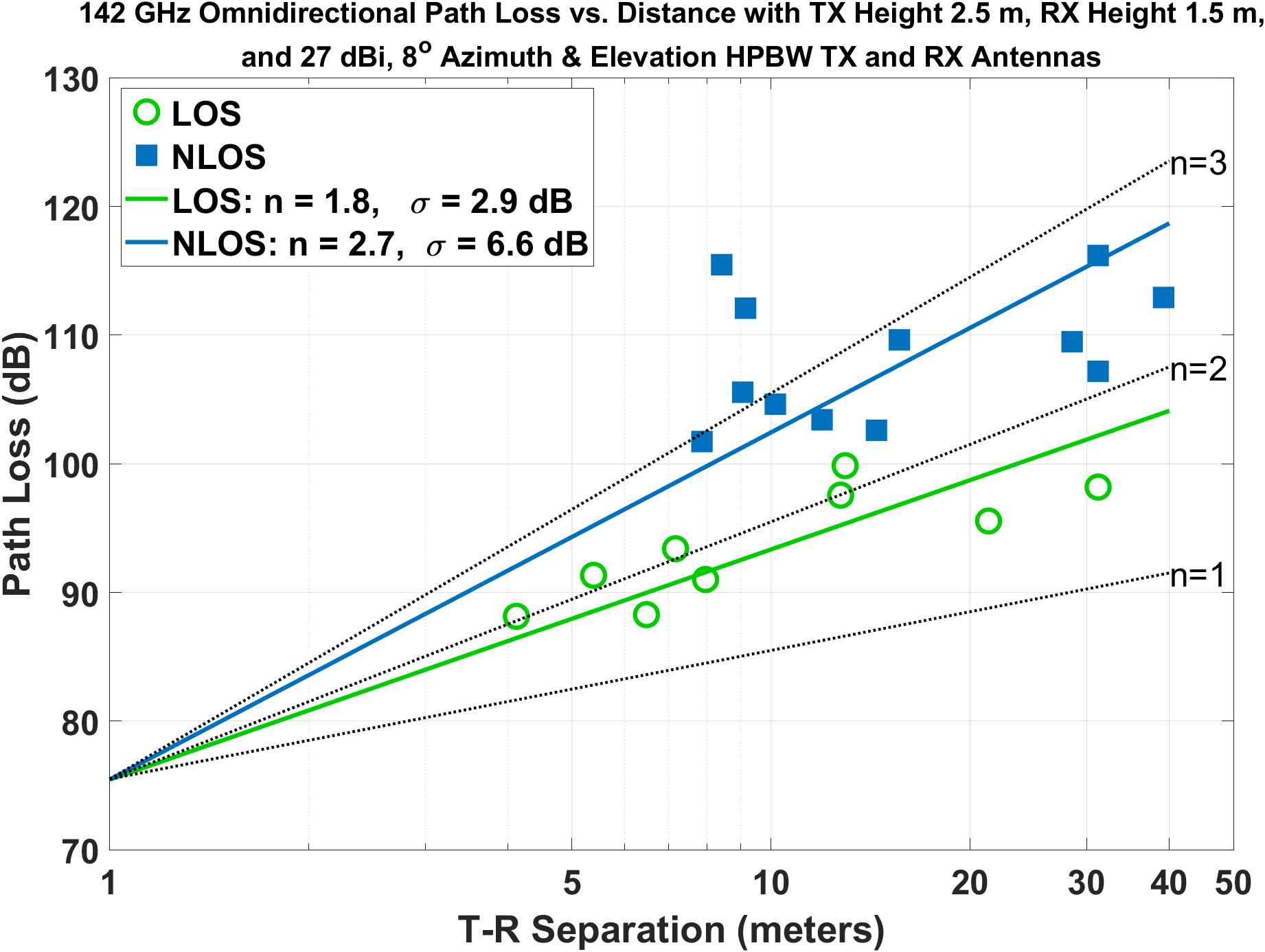}
		\caption{142 GHz indoor omnidirectional CI path loss model and data.}
		\label{fig:omni_pl_140}
	\end{subfigure}
	\caption{28 GHz and 142 GHz indoor omnidirectional CI path loss models and scatter plots with TX antenna height of 2.5 m and RX antenna height of 1.5 m for V-V polarization.}
	\label{fig:omni_pl}
\end{figure}

Even though the directional path loss model will be widely used in future wireless system deployment, the omnidirectional path loss model is fundamental and serves as a reference model in various standard documents \cite{80211ay16, 3GPP38901r16}. In Fig. \ref{fig:omni_pl}, we present the omnidirectional path loss data and the fitted CI path loss model. The omnidirectional path loss is synthesized from received powers from all directions measured in the 3-D space \cite{Sun15synthesize}. Fig. \ref{fig:omni_pl} marks LOS and NLOS scenarios in green and blue, respectively. The LOS PLEs at both frequencies are lower than 2.0, where 28 GHz shows a surprisingly low PLE of 1.2, which can be attributed to the waveguide effect in some corridor measurement locations. Note that both 28 and 142 GHz have a comparable PLE of about 2.7 in the NLOS environment, indicating that the signal power drops equally versus distances after the first meter in the mmWave band of 28 GHz and the sub-THz band of 140 GHz \cite{Mac15access,Xing21b}.

\section{3-D Spatial Statistical Channel Model} \label{sec:method}
A received signal can be viewed as a superposition of multiple replicas of the transmitted signal with different delays and angles for any wireless propagation channel \cite{Rap02textbook}. An extended S-V channel model \cite{Saleh87} was commonly used to represent the double directional channel in the 3-D space \cite{Wu17tap,3GPP38901r16}. MPCs were observed to arrive in clusters in delay and angular domains from 28 GHz and 140 GHz indoor channel measurements, which agreed with many early works \cite{Saleh87,Wu17tap}. Current standards document such as the 3GPP TR 38.901 channel model defined a cluster as a group of MPCs closely spaced in the joint temporal-spatial domain, where each cluster represented a reflector or a scatterer in the environment \cite{Wu17tap,80211ad10,3GPP38901r16}. 


We observed in the measurements that MPCs traveling close in time may arrive from very different directions due to the symmetric structure of the environment like hallways \cite{Mac17sounder,Rap13access}. Conversely, MPCs arriving from a similar direction may have very different propagation times. A time cluster spatial lobe (TCSL) approach was introduced to characterize temporal and angular domains separately \cite{Samimi16mtt}. A time cluster (TC) comprises MPCs traveling close in time and arriving from potentially different directions. A spatial lobe (SL) represents a main direction of arrival or departure where MPCs can arrive over hundreds of nanoseconds \cite{Samimi16mtt}. 

Both modeling methodologies are valid, where the 3GPP model is more widely used and the NYUSIM model using TCSL has a more straightforward and physically-based structure \cite{Polese17jsac,Barati15twc}. Performance evaluation with respect to spectrum efficiency, coverage, and hardware/signal processing requirements between the 3GPP and NYUSIM channel models were provided in \cite{Sun18tvt}. 

The cluster-based omnidirectional CIR $h_{\text{omni}}(t,\overrightarrow{\Theta},\overrightarrow{\Phi})$ is given by: 
\begin{equation}
	\label{eq:omni_cir}
	\begin{split}
		h_{\text{omni}}(t,\overrightarrow{\Theta},\overrightarrow{\Phi}) = &\sum_{n=1}^{N}\sum_{m=1}^{M_n}a_{m,n}e^{j\varphi_{m,n}}\cdot\delta(t-\tau_{m,n})\\
		&\cdot\delta(\overrightarrow{\Theta}-\overrightarrow{\Theta_{m,n}}) \cdot\delta(\overrightarrow{\Phi}-\overrightarrow{\Phi_{m,n}}),
	\end{split}
\end{equation}
where $t$ is the absolute propagation time, $\overrightarrow{\Theta}=(\phi_{\textup{AOD}},\theta_{\textup{ZOD}})$ is the AOD vector, and $\overrightarrow{\Phi}=(\phi_{\textup{AOA}},\theta_{\textup{ZOA}})$ is the AOA vector. $N$ and $M_n$ denote the number of TCs and the number of subpaths within each TC, respectively. For the $m$th subpath in the $n$th TC, $a_{m,n}$, $\varphi_{m,n}$, $\tau_{m,n}$, $\overrightarrow{\Theta_{m,n}}$, and $\overrightarrow{\Phi_{m,n}}$ represent the magnitude, phase, absolute time delay, AOD vector and AOA vector, respectively. Note that MPC and subpath are used interchangeably. The PDP and APS can be obtained by integrating the square of the CIR in space and time domains, respectively. 

The PDP and APS can be easily partitioned based on TCs and SLs, respectively (see Fig. 10 and Fig. 11 in \cite{Samimi16mtt}). The partition in the time domain is realized by defining a minimum inter-cluster time void interval (MTI). Two sequentially recorded MPCs belong to two distinct TCs if the difference of the excess time delays of these two MPCs is beyond MTI. These two MPCs are considered as the last MPC of the former TC and the first MPC of the latter TC, respectively. For example, 25 ns was used as MTI for an outdoor urban microcell (UMi) environment \cite{Samimi16mtt}, while 6 ns is used as MTI in this paper for an indoor office (InO) environment since the width of a typical hallway in the measured indoor office environment is about 1.8 m (i.e., $\sim$6 ns propagation delay). 


The partition in the space domain is realized by defining a spatial lobe threshold (SLT) \cite{Samimi16mtt}. The angular resolution of the measured APS depends on the antenna HPBW (30\degree~and 8\degree~for 28 GHz and 140 GHz, respectively). A linear interpolation of the directional received powers in azimuth and elevation planes with 1\degree~resolution was used to reconstruct the 3-D spatial distribution of the received power. A power segment is generated for every 1\degree~direction in the 3-D space. Neighboring power segments above the SLT form an SL. The SLT was -15 dB below the maximum directional power in the APS.


As defined in \cite{Samimi16mtt}, the primary statistics such as the number of TCs and SLs, cluster delays, and cluster powers are used in the channel generation procedure given in Section \ref{sec:stat}. The secondary statistics such as RMS DS and RMS AS are not required in the channel generation but necessary in the channel validation. The presented channel model will be validated in Section \ref{sec:simulation} by showing that the simulated and measured secondary statistics yield good agreements.

\section{Statistics of Channel Generation Parameters} \label{sec:stat}
As described in Section \ref{sec:method}, temporal and spatial channel parameters are extracted from the measured PDP and APS. Temporal parameters are the number of TCs ($N$) and SPs in a TC ($M_n$), TC excess delay ($\tau_n$) and intra-cluster subpath excess delay ($\rho_{m,n}$), TC power ($P_n$) and subpath power ($\Pi_{m,n}$). Spatial parameters are the number of SLs ($L$), the mean azimuth and elevation angle of an SL ($\phi$ and $\theta$), and the azimuth and elevation angular offset of a subpath ($\Delta \phi$ and $\Delta \theta$) with respect to the mean angle of the SL.

Since the 140 GHz measurement locations is a subset of the 28 GHz measurement locations, the \textit{28 GHz common set} was created out of the \textit{28 GHz all set} to have a fair comparison with the 140 GHz dataset (referred to the \textit{140 GHz common set} below). The 28 GHz common set and the 140 GHz common set have identical TX-RX location pairs. Table \ref{tab:model} presents the channel parameters required for channel generation procedure. Table \ref{tab:values} provides statistics of channel parameters derived from the 28 GHz all set, 28 GHz common set, and 140 GHz common set, for LOS and NLOS scenarios.
\begin{table*}[]
	\centering
	\caption{\textsc{Input Parameters for channel coefficient generation procedure}}
	\label{tab:model}
	\begin{tabular}{|c|c|c|c|c|}
		\hline
		\textbf{Step Index} &\textbf{Channel Parameters}                                            & \textbf{28 - 73 GHz UMi \cite{Samimi16mtt}}                                                                                                                                                                                                           & \multicolumn{2}{c|}{\textbf{28 - 140  GHz InO}}                                                                                                                                                                       \\ \hline
		\textbf{Step 1} &\textbf{\# Time clusters $N$}                                          & $N \sim $ \textup{DU}$(1, N_c)$                                                                                                                                                                                                           & \multicolumn{2}{c|}{$N \sim \textup{Poisson}(\lambda_c)$}                                                                                                                                                                                           \\ \hline
		\textbf{Step 2} &\textbf{\# Cluster subpaths $M_n$}                                     & $M_n \sim $ DU$(1, M_s)$                                                                                                                                                                                                         & \multicolumn{2}{c|}{$M_n \sim (1-\beta)\delta(M_n)+\textup{DE}(\mu_s)$        }                                                                                                                                                  \\ \hline
		\textbf{Step 3} &\textbf{Cluster delay $\tau_n$ (ns)}                                   & \multicolumn{3}{c|}{$\!\begin{aligned}\tau_n'' &\sim \textup{Exp}(\mu_{\tau})  \:\text{or}\: \textup{Logn}(\mu_{\tau},\sigma_{\tau}) \\ \Delta \tau_n&=\textup{sort}(\tau_n'')-\textup{min}(\tau_n'')\\ \tau_n &= \begin{cases} 0 & ,n=1\\ \tau_{n-1}+\rho_{M_{n-1},n-1}+\Delta\tau_n+\textup{MTI} &,n=2,...N \end{cases}\end{aligned}$}                                                                                                                                                                                                                                                                             \\ \hline
		\textbf{Step 4} &\textbf{Intra-cluster delay $\rho_{m,n}$ (ns)}                         & {$\!\begin{aligned}\rho_{m,n}=\left[\frac{1}{B_{bb}}\times (m-1)\right]^{1+X_n},\\  m = 1,2,...,M_n, \: n=1,2,...,N\end{aligned}$}                                               & \multicolumn{2}{c|}{$\rho_{m,n}\sim $ Exp$(\mu_{\rho})$}                                                                                                                                                                                                                                                                       \\ \hline
		\textbf{Step 5} &\textbf{Cluster power $P_n$ (mW)}                                 & \multicolumn{3}{c|}{$\!\begin{aligned}	P'_n &= \bar{P}_0 e^{-\frac{\tau_n}{\Gamma}}10^{\frac{Z_n}{10}}, \label{eq:cp1}\\
				P_n &= \frac{P'_n}{\sum_{k=1}^N P'_k} \times P_r [mW], \label{eq:cp2}\\
				Z_n&\sim \mathcal{N}(0, \sigma_Z), \;n = 1,2,...,N \label{eq:cp3}\end{aligned}$} \\ \hline
		\textbf{Step 6} &\textbf{Subpath power $\Pi_{m,n}$(mW)}                                      & \multicolumn{3}{c|}{$\!\begin{aligned}
				\Pi'_{m,n} &= \bar{\Pi}_0 e^{-\frac{\rho_{m,n}}{\gamma}}10^{\frac{U_{m,n}}{10}}, \label{eq:sp1}\\
				\Pi_{m,n} &= \frac{\Pi'_{m,n}}{\sum_{k=1}^{M_n} \Pi'_{k,n}} \times P_n [mW], \label{eq:sp2}\\
				U_{m,n}&\sim \mathcal{N}(0, \sigma_U), \;m = 1,2,...,M_n \label{eq:sp3}
			\end{aligned}$}                                                                                                                                                                                                                             \\ \hline
		\textbf{Step 7} &\textbf{Subpath phase $\varphi$ (rad)}                                                &\multicolumn{3}{c|}{Uniform(0, 2$\pi$)}        \\ \hline
		\textbf{Step 8} &\textbf{\# Spatial lobes $L$}                                          & {\begin{tabular}[c]{@{}c@{}} $L_{\textup{AOD}}\sim \text{min}\{L_{\textup{max}},\max\{1,\textup{Poisson}(\mu_{\textup{AOD}})\}\}$\\$L_{\textup{AOA}}\sim \text{min}\{L_{\textup{max}},\max\{1,\textup{Poisson}(\mu_{\textup{AOA}})\}\}$
		\end{tabular}}                                              & \multicolumn{2}{c|}{\begin{tabular}[c]{@{}c@{}} $L_{\textup{AOD}}\sim \text{DU}(1,L_{\textup{AOD,max}})$\\$L_{\textup{AOA}}\sim \text{DU}(1,L_{\textup{AOA,max}})$
		\end{tabular}}          \\ \hline
		\textbf{Step 9} &\textbf{Spatial lobe mean angle $\phi_i,\theta_i$ (\degree)}                                                 &\multicolumn{3}{c|}{\begin{tabular}[c]{@{}c@{}} $\phi_i\sim U(\phi_{\textup{min}},\phi_{\textup{max}}),\phi_{\textup{min}}=\frac{360(i-1)}{L},\phi_{\textup{max}}=\frac{360i}{L},i=1,2,...,L$\\$\theta_i\sim \mathcal{N}(\mu_{l},\sigma_{l})$
		\end{tabular}}                                                                                                                                          \\ \hline
		\textbf{Step 10} &\textbf{Subpath angle offset $\Delta\phi_i,\Delta\theta_i$ w.r.t $\phi_i,\theta_i$ (\degree)} &\multicolumn{3}{c|}{$\!\begin{aligned}
				i\sim\textup{DU}[1,L_{\textup{AOD}}]&,j\sim\textup{DU}[1,L_{\textup{AOA}}]\\
				(\Delta\phi_i)_{m,n,\textup{AOD}}&\sim \mathcal{N}(0,\sigma_{\phi,\textup{AOD}})\\
				(\Delta\theta_i)_{m,n,\textup{ZOD}}&\sim \mathcal{N}(0,\sigma_{\theta,\textup{ZOD}}) \\
				(\Delta\phi_j)_{m,n,\textup{AOA}}&\sim \mathcal{N}(0,\sigma_{\phi,\textup{AOA}})\\
				(\Delta\theta_j)_{m,n,\textup{ZOA}}&\sim \mathcal{N}(0,\sigma_{\theta,\textup{ZOA}})\\
			\end{aligned}$}                                                                                                 \\ \hline
	\end{tabular}
\end{table*}

\subsection{Temporal Channel Parameters}
\subsubsection{The Number of Time Clusters}
\begin{figure}
	\centering
	\begin{subfigure}[b]{.45\textwidth}
		\centering
		\includegraphics[width=0.9\linewidth]{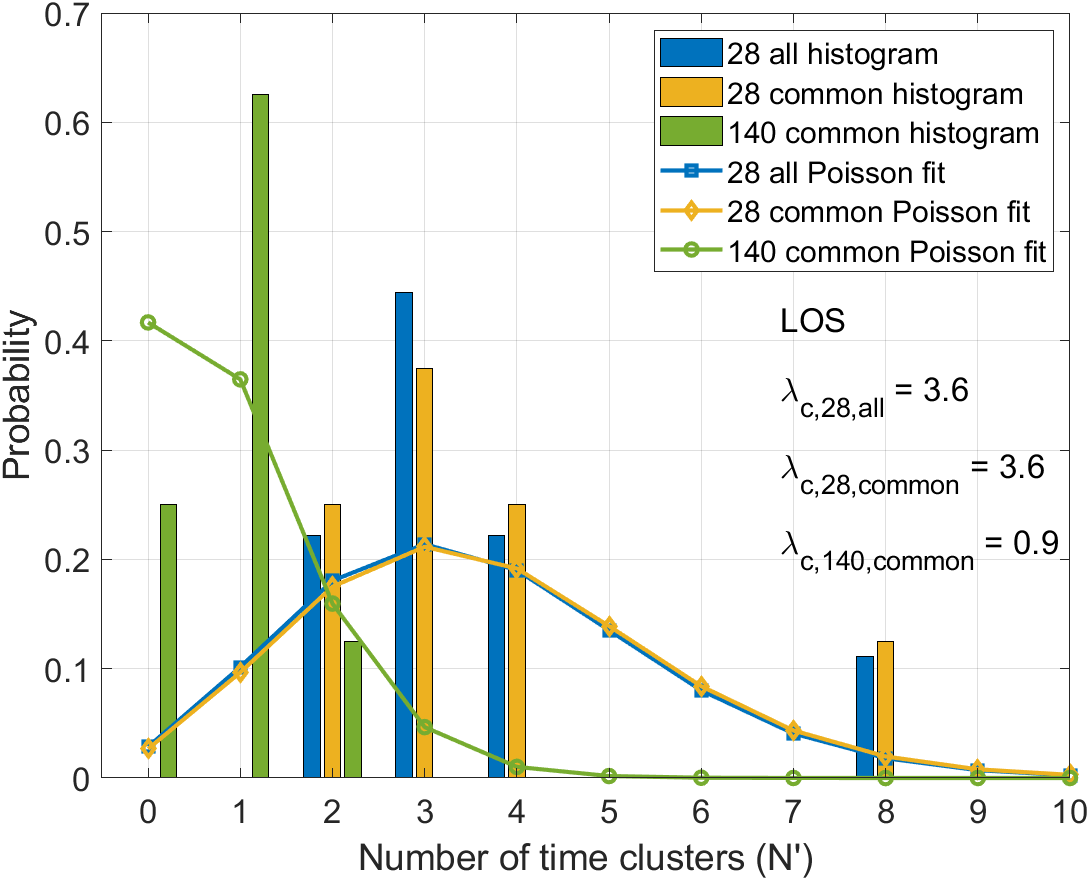}
		\caption{The number of TCs in the NLOS scenario.}
		\label{fig:num_tc_nlos_28_140}
	\end{subfigure}
	\hfill
	\begin{subfigure}[b]{.45\textwidth}
		\centering
		\includegraphics[width=0.9\linewidth]{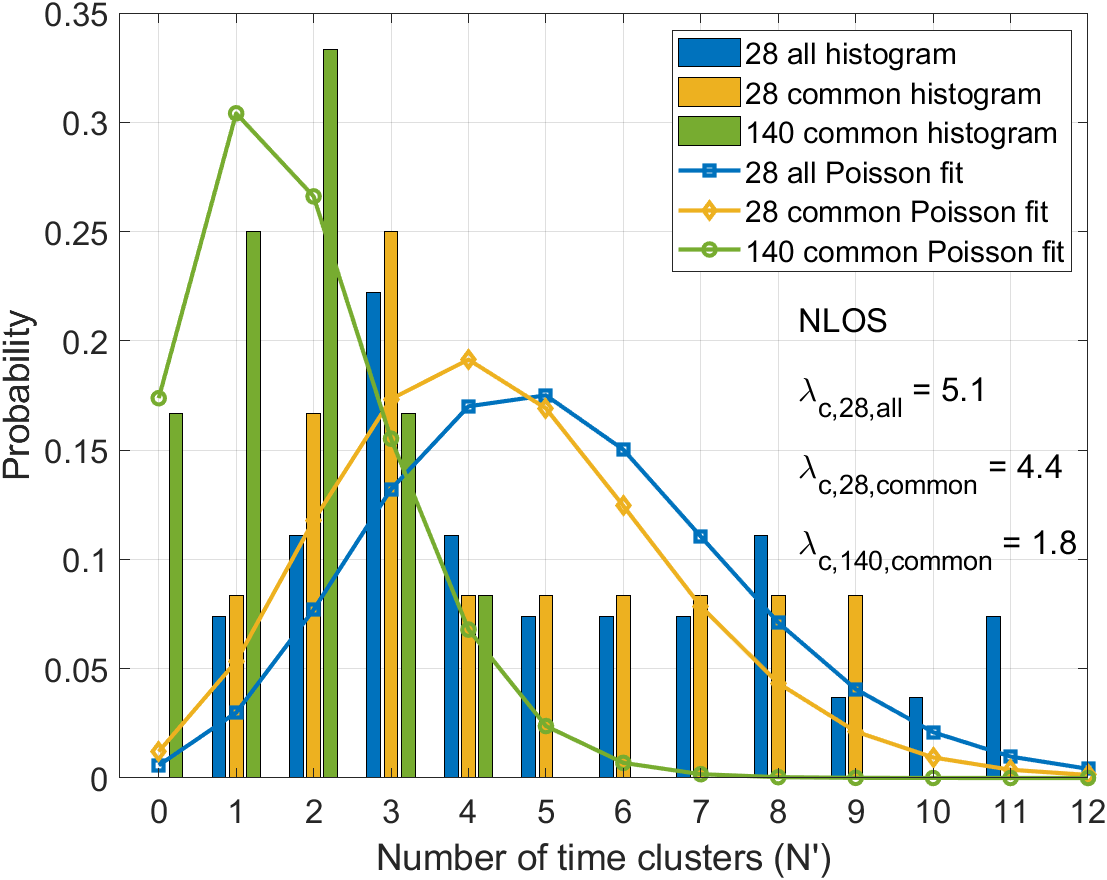}
		\caption{The number of TCs in the LOS scenario.}
		\label{fig:num_tc_los_28_140}
	\end{subfigure}
	\caption{Histograms and Poisson distribution fittings of the number of TCs of 28 GHz all set, 28 GHz common set, and 140 GHz common set in the (a) NLOS scenario (b) LOS scenario.}
	\label{fig:num_tc_all}
\end{figure}
TCs are obtained by partitioning the measured PDPs based on the MTI. In Fig. \ref{fig:num_tc_all}, the empirical histograms of the number of TCs ($N$) of three datasets (i.e., 28 GHz all set, 28 GHz common set, and 140 GHz common set) for the LOS and NLOS scenarios with a 6 ns MTI are shown to be well fitted by the Poisson distribution. Since the Poisson distribution starts from zero while the number of TCs is at least one. Thus, $N'=N-1$ is used for distribution fitting, and the maximum likelihood estimator (MLE) of the parameter $\lambda_c$ of the Poisson distribution is the sample mean of $N'$. The simulated number of TCs from the Poisson distribution is added by one to obtain the actual number of TCs, which is given by:
\begin{equation}
	\begin{split}
		P(N'=k)& = \frac{\lambda_c^k}{k!}e^{-\lambda_c},\quad\quad k=0,1,2,...,\\
		N &= N'+1.
	\end{split}
\end{equation}
The 28 GHz channel has about three more TCs than the 140 GHz channel in both NLOS and LOS scenarios, which can be attributed to the higher partition loss at 140 GHz (e.g., 4-8 dB higher than 28 GHz for different materials \cite{Xing19globecom}). The channel sparsity at 140 GHz should be considered in the channel estimation and beamforming algorithms for sub-THz frequencies. The NLOS scenario has about one more TC than the LOS scenario. Note that the Poisson distribution of the number of TCs for the indoor scenario is different from the uniform distribution used for the outdoor scenario, as given in Table \ref{tab:model} \cite{Samimi16mtt}. 

\subsubsection{Number of Cluster Subpaths}

\begin{figure}
	\centering
	\begin{subfigure}[b]{.45\textwidth}
		\centering
		\includegraphics[width=0.9\linewidth]{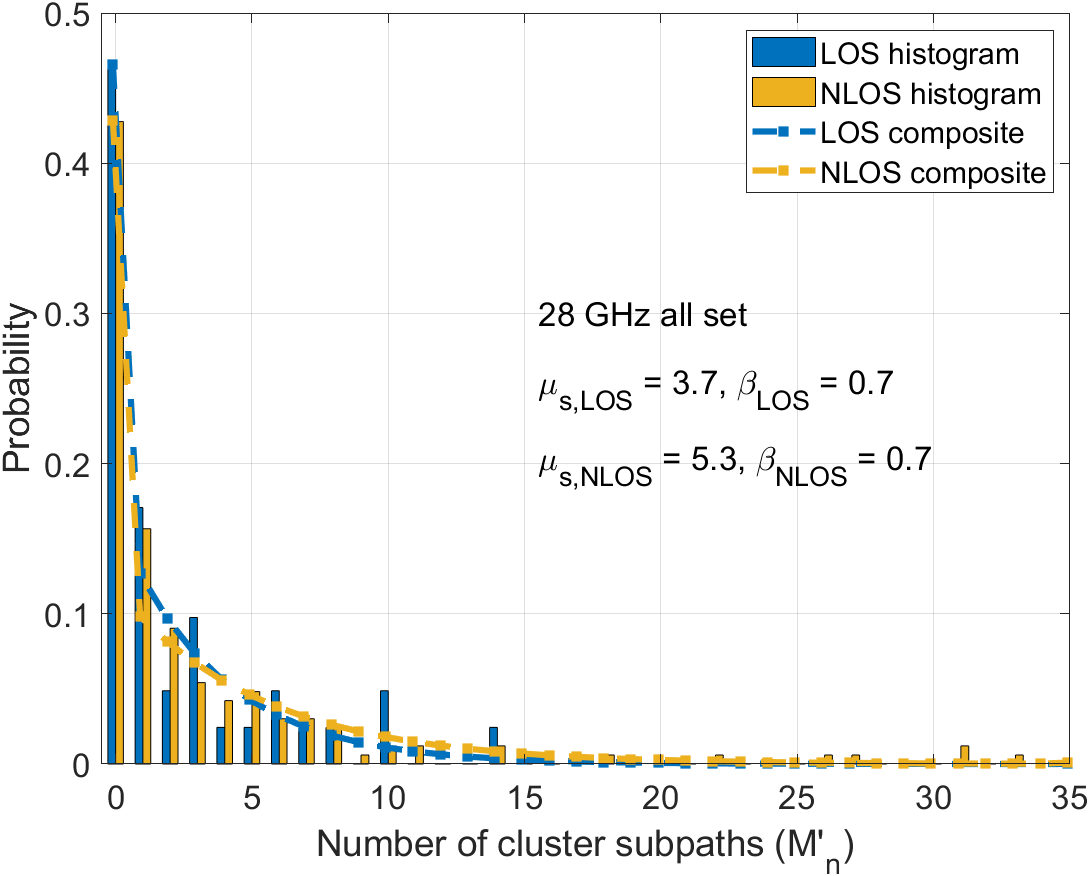}
		\caption{The number of subpaths for 28 GHz all set.}
		\label{fig:num_sp_28}
	\end{subfigure}
	\hfill
	\begin{subfigure}[b]{.45\textwidth}
		\centering
		\includegraphics[width=0.9\linewidth]{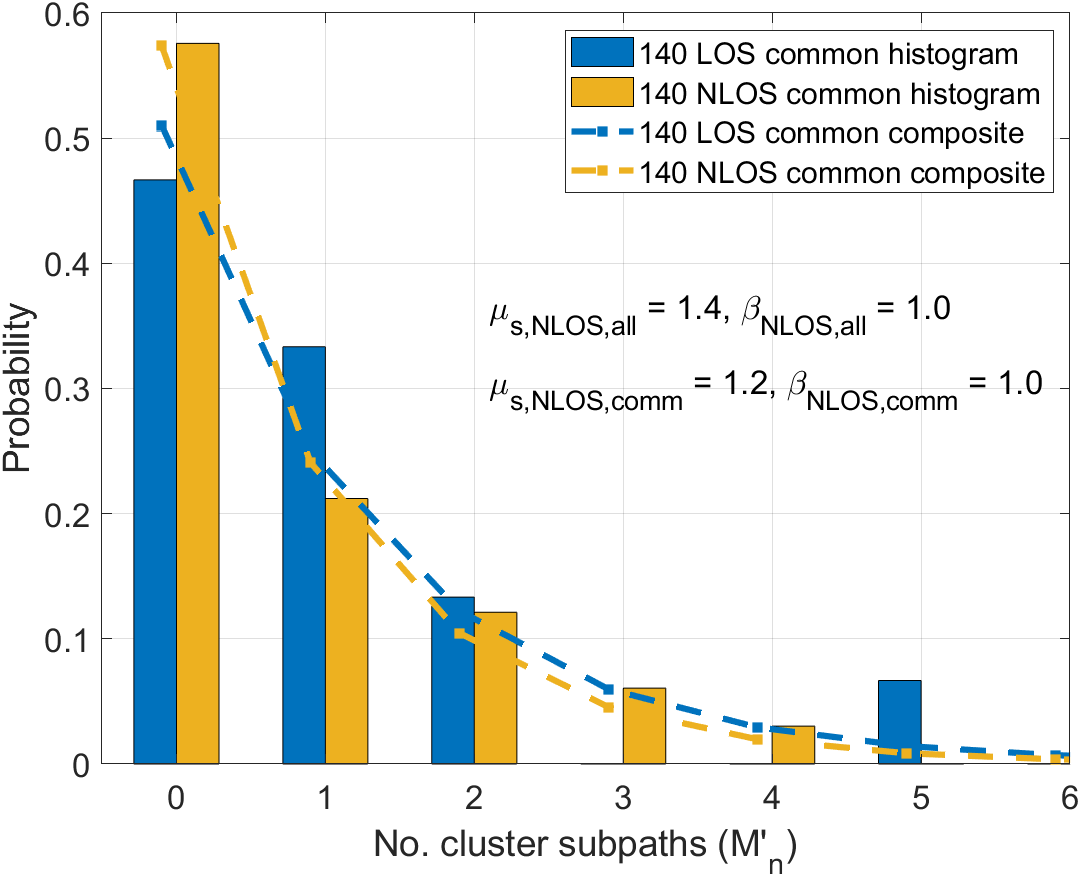}
		\caption{The number of subpaths for 140 GHz common set.}
		\label{fig:num_sp_140}
	\end{subfigure}
	\caption{Histograms and composite distribution fittings of the number of subpaths of (a) 28 GHz all set and (b) 140 GHz common set.}
	\label{fig:num_sp_all}
\end{figure}



The number of cluster subpaths $M_n$ is negatively correlated to the number of TCs depending on the MTI. A larger MTI causes fewer TCs and more subpaths within each TC, and vice versa. Fig. \ref{fig:num_sp_all} presents the empirical histograms for three datasets, indicating that the number of cluster subpaths is close to exponentially distributed. The exponential distribution is continuous and starts from zero, while the value of the number of subpaths is discrete and starts from one. Thus, a discrete exponential (DE) distribution is applied to fit the empirical histogram of $M_n'=M_n-1$. Fig. \ref{fig:num_sp_28} shows that about half of the measured TCs only have one subpath at 28 GHz, making a simple DE distribution unsuitable. We proposed a composite distribution with a $\delta$-function at $M_n'=0$ and a DE distribution, which is given by
\begin{equation}
	\begin{split}
		P_{M_n'}(k) = (1-\beta)\delta(k)&+\beta\int_{k}^{k+1}\frac{1}{\mu_s}e^{-\frac{x}{\mu_s}}dx,\\
		&k=0,1,2,...,
	\end{split}
\end{equation}
where $\mu_s$ is the mean of the DE distribution, and $\beta$ is the weight of the DE distribution in the composite distribution. By maximizing the joint probability mass function (PMF) of all data samples over $\beta$ and $\mu_s$ simultaneously, the MLE of $\mu_s$ and $\beta$ is 5.3 and 0.7 for 28 GHz NLOS all set, respectively. The identical composite distribution for 28 GHz LOS all set shows that $\mu_s=3.7$, $\beta=0.7$, suggesting that the NLOS scenario forms larger clusters than the LOS scenario. The composite distribution yields a good agreement with empirical histograms of 28 GHz LOS and NLOS scenarios. Moreover, the large TCs with more than 25 subpaths were mainly from locations in the corridor environment (e.g., TX4 and RX15).

Fig. \ref{fig:num_sp_140} shows that the a simple DE distribution matches the empirical histogram of 140 GHz LOS and NLOS common sets since the optimal $\beta_{\textup{NLOS}}=1$. The mean number of $M_n'$ for both LOS and NLOS scenarios is about 1, suggesting that LOS and NLOS scenarios have similar sizes of clusters which contain about two subpaths on average. The clusters at 140 GHz are much smaller than the clusters at 28 GHz, indicating that the 140 GHz channel is much sparser than the channel at 28 GHz. The detailed comparison of channel parameters between 28 GHz and 140 GHz common sets can be found in Table \ref{tab:values}.

\subsubsection{Inter-cluster Excess Delay}
\begin{figure}[]
	\centering
	\includegraphics[width=0.9\linewidth]{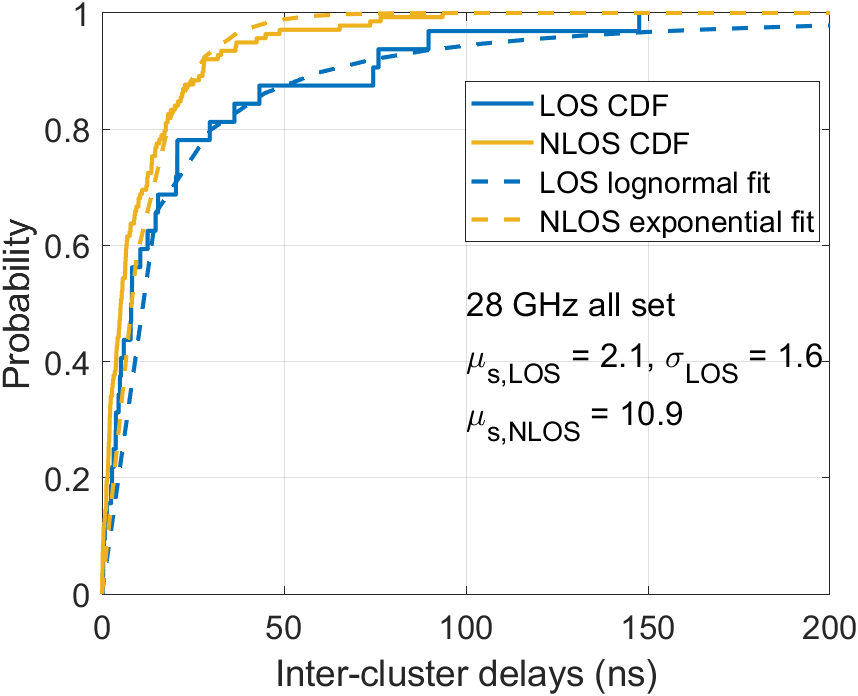}
	\caption{Inter-cluster delays for 28 GHz LOS and NLOS all set.}
	\label{fig:cd_28}
\end{figure}

The cluster excess delay $\tau_n$ is defined as the time difference between the first arriving subpath in the PDP and the first arriving subpath in a cluster, as given in Table \ref{tab:model}, where $\tau_{n-1}$ is the cluster excess delay of the former cluster, $\rho_{M_{n-1},n-1}$ is the intra-cluster excess delay of the last subpath in the former cluster. $\Delta \tau_n$ is the inter-cluster excess delay without MTI (i.e., 6 ns). The empirical cumulative distribution function (CDF) of the inter-cluster delay for LOS and NLOS scenarios of 28 GHz all set are shown in Fig. \ref{fig:cd_28}. An exponential distribution with the mean 10.9 ns fits the 28 GHz NLOS scenario, while a lognormal distribution with the mean 2.1 ns and standard deviation 1.6 ns fits the 28 GHz LOS scenario well since a few clusters with long cluster delays were observed in the LOS corridor environment. 

Inter-cluster delays at 140 GHz can be well fitted using an exponential distribution for both LOS and NLOS scenarios, where the mean values are 14.6 ns and 21.0 ns, respectively. The distributions for the 28 GHz and 140 GHz LOS scenarios are different (lognormal and exponential), likely due to the higher partition loss and the smaller measurable range at 140 GHz. In addition, clusters with large inter-cluster delays were mainly observed in the corridor environment due to the waveguide effect, indicating that the corridor scenario may be considered a distinct indoor scenario and requires more channel measurements for accurate characterization.  

\subsubsection{Intra-cluster Excess Delay}
\begin{figure}[h!]
	\centering
	\includegraphics[width=0.9\linewidth]{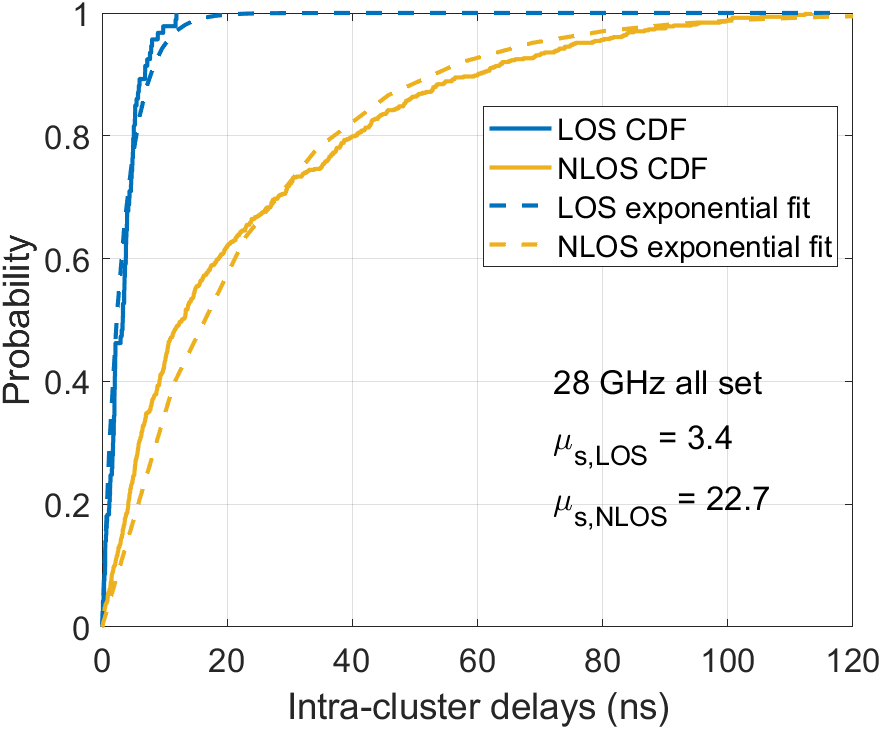}
	\caption{Intra-cluster delay for 28 GHz LOS and NLOS all set. }
	\label{fig:sd_28}
\end{figure}

The intra-cluster excess delay is defined as the time difference between the first arriving subpath and the targeted arriving subpath within the same TC. As shown in Fig. \ref{fig:sd_28}, an exponential distribution shows a good agreement with the empirical CDF for 28 GHz LOS and NLOS scenarios, where the mean intra-cluster excess delay is 3.4 ns and 22.7 ns for 28 GHz LOS and NLOS all set, suggesting a larger intra-cluster delay is usually observed in the NLOS scenario. 
\subsubsection{Cluster Power and Subpath Power}
\begin{figure}[h!]
	\centering
	\includegraphics[width=0.9\linewidth]{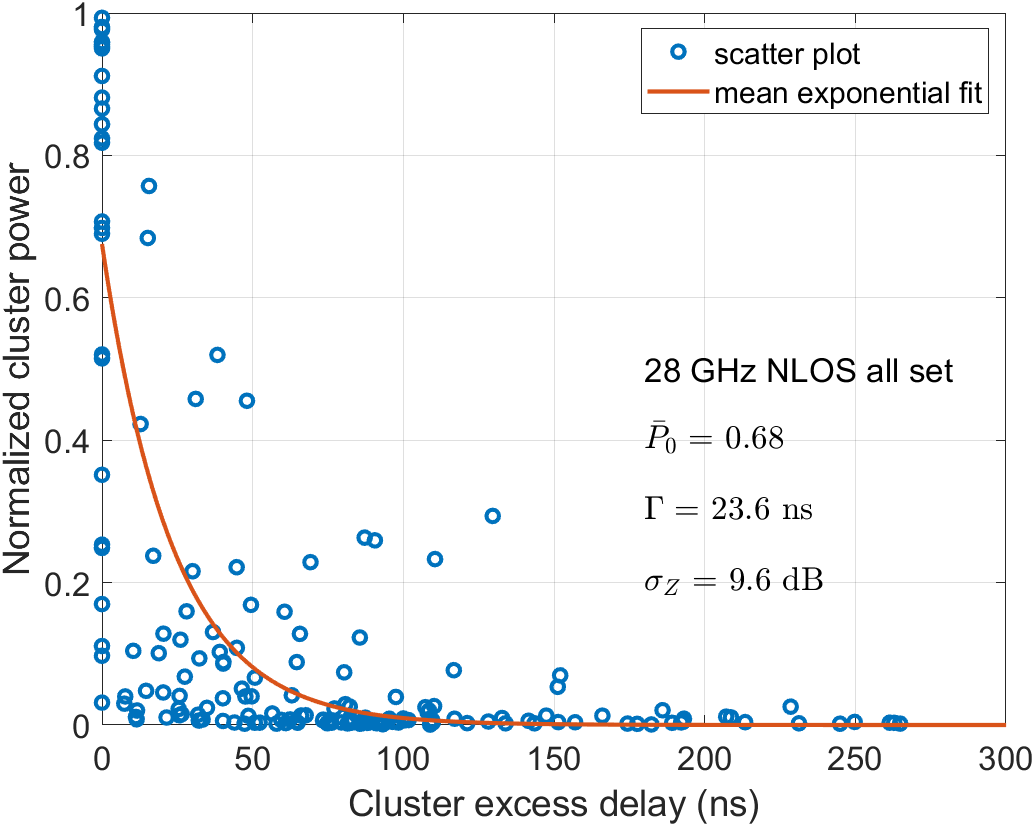}
	\caption{Normalized cluster powers for 28 GHz NLOS all set. }
	\label{fig:cp_28_nlos}
\end{figure}

Cluster power is defined as the sum of the subpath powers in the cluster, and the normalized cluster power over the total received power in the PDP can be well modeled by an exponentially decaying function of cluster excess delay with a lognormal-distributed shadowing term, as given in Table \ref{tab:model}. $\bar{P}_0$ is the mean power in the first arriving TC, $\Gamma$ is the cluster decay time constant, and $Z_n$ is a lognormal distributed (normal in dB scale) shadowing term for the cluster power with zero-dB mean and standard deviation $\sigma_Z$. $P_n$ represents the cluster power so that the sum of $P_n$ is equal to the total omnidirectional received power $P_r$. The normalized cluster powers measured in the 28 GHz NLOS scenario is shown in Fig. \ref{fig:cp_28_nlos}, where $P_0$ is 0.68, and $\Gamma$ is 23.6 ns, indicating that the expected first cluster occupies about 68\% of the total received power and the expected cluster power is less than 34\% of the total received power when the cluster excess time delay is over 23.6 ns. 

\begin{figure}[h!]
	\centering
	\includegraphics[width=0.9\linewidth]{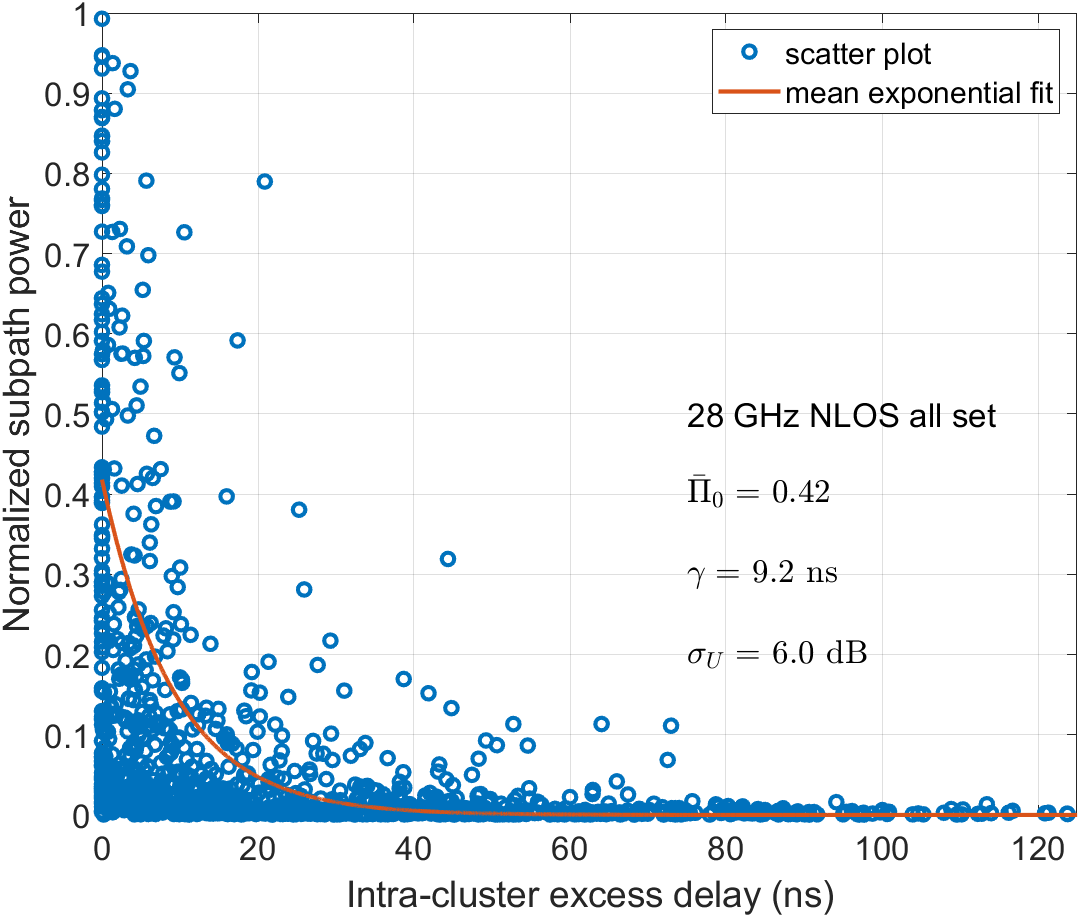}
	\caption{Normalized subpath powers for 28 GHz NLOS all set. }
	\label{fig:sp_28_nlos}
\end{figure}

Similarly, the normalized subpath power over the cluster power can be modeled as an exponentially decaying function over the intra-cluster excess delay, as given in Table \ref{tab:model}. $\bar{\Pi}_0$ is the mean power in the first arriving subpath in a TC. $\gamma$ is the subpath decay time constant, and $U_{m,n}$ is a lognormal distributed shadowing term for the subpath power with zero-dB mean and standard deviation $\sigma_U$. Fig. \ref{fig:sp_28_nlos} shows that the first subpath in the cluster is about 42\% of the cluster power on average, suggesting a relatively large RMS intra-cluster DS. The expected subpath power is less than 21\% of the cluster power when the intra-cluster excess time delay is over 9.2 ns.

\subsection{Spatial Channel Parameters}
\subsubsection{The Number of Spatial Lobes}
\begin{figure}[h!]
	\centering
	\includegraphics[width=0.9\linewidth]{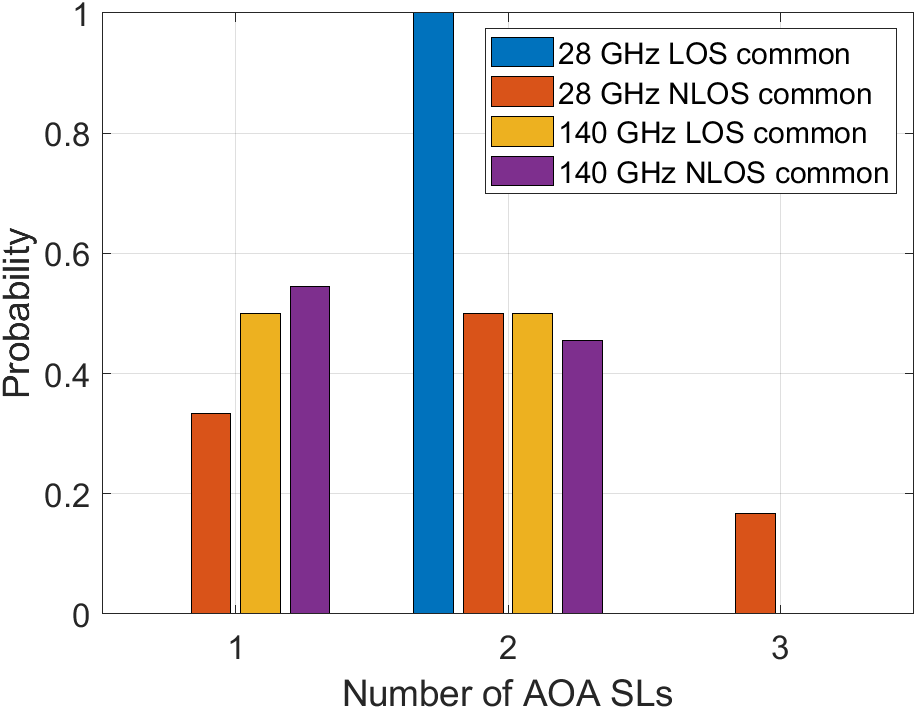}
	\caption{\textcolor{black}{Histograms of the number of AOA spatial lobes for 28 GHz and 140 GHz common sets.} }
	\label{fig:nsl}
\end{figure}

An SL represents a main direction of arrival or departure. The angular resolution of the measured APS depends on the antenna HPBW, which are 30\degree~and 8\degree~in 28 and 140 GHz measurements, respectively. A linear interpolation of the measured directional powers with 1\degree~angular resolution in the azimuth and elevation planes was implemented to model the 3-D spatial distribution of the received power. The SLT is -15 dB below the peak power. Measurement results show that there are at most two main directions of arrival in the azimuth plane, except that there are a few NLOS locations measured at 28 GHz which observed three main directions of arrival, as shown in Fig. \ref{fig:nsl}. Thus, a simple DU distribution is used to characterize the number of spatial lobes, which is given in Table \ref{tab:model}. 

\subsubsection{Mean Direction of Spatial Lobes}
Each SL has a mean direction in the azimuth and elevation planes. A simple partition can be applied to generate the azimuth mean direction of an SL by equally dividing the azimuth plane into several sectors, each of which corresponds to an SL. The elevation mean direction of an SL is modeled as a normal random variable $N(\mu_l,\sigma_l)$, as given in Table \ref{tab:model}, where $\theta_i$ is defined with respect to the horizontal plane. Considering that the TX height is usually higher than the RX height in a downlink of base station to mobile device setting, $\mu_l$ of ZOD is typically negative, and $\mu_l$ of ZOA is typically positive, which represents that ZOD and ZOA are below and above the horizon, respectively. 

\subsubsection{Subpath Angular Offset}
For each spatial lobe, the RMS lobe AS is extracted from the partitioned AOA and AOD APS. A generated subpath is randomly assigned to one of the generated SLs, and the angles of this subpath (i.e., AOD, ZOD, AOA, and ZOA) are calculated by adding angular offsets with respect to the mean angle of the SL. The angular offset follows a normal distribution with zero mean and a standard deviation of the median of the measured RMS lobe AS, as given in Table \ref{tab:model}. Such angular offset generation deviates from the 3GPP TR 38.901 channel model where angular offsets of 20 MPCs in a cluster are constant \cite{3GPP38901r16}. 

\subsection{Discussions}
Each temporal and spatial channel parameter discussed above is generally fitted well by an identical distribution for 28 and 140 GHz, but the values of each parameter for these two frequencies are quite different. The channel at 140 GHz has fewer time clusters and fewer subpaths within each cluster than the channel at 28 GHz. Greater partition loss and higher path loss in the first meter of propagation distance at 140 GHz cause a smaller signal propagation range (the difference of maximum measurable path loss between two frequencies has been considered); thus, some of RX locations which could receive signals at 28 GHz were in outage at 140 GHz.
\begin{table*}[]
	\centering
	\caption{\textsc{Values of required parameters in the channel generation procedure derived from 28 GHz all set, 28 GHz common set, and 140 GHz common set for LOS and NLOS scenarios.}}
	\label{tab:values}
	\begin{tabular}{|c|c|c|c|c|c|c|}
		\hline
		\textbf{Input Parameters}                                                          & \textbf{28 LOS all}   & \textbf{28 NLOS all } & \textbf{28 LOS common}   & \textbf{28 NLOS common} & \textbf{140 LOS common}  & \textbf{140 NLOS common} \\ \hline
		\textbf{$\lambda_c$}                                                               & 3.6                     & 5.1                    
		& 3.6                    & 4.4				& 0.9                     & 1.8                  \\ \hline
		\textbf{$\beta_s$}                                                                 & 0.7                   & 0.7                  & 0.7                   & 0.7		 			& 1.0                   & 1.0                   \\ \hline
		\textbf{$\mu_s$}                                                                   & 3.7                   & 5.3                   & 3.4                   & 4.6  				& 1.4                   & 1.2                   \\ \hline
		\textbf{$\mu_{\tau}[\textup{ns}]$}                                                 & logn(2.1, 1.6)        & 10.9                  & logn(1.9,1.6)        & 9.8 			    & 14.6                  & 21.0                   \\ \hline
		\textbf{$\mu_{\rho}[\textup{ns}]$}                                                 & 3.4                   & 22.7                 & 3.4                   & 14.2     			& 1.1                   & 2.7                    \\ \hline
		\textbf{$\Gamma[\textup{ns}],\sigma_{Z}[\textup{dB}]$}                             & 20.7, 15.4             & 23.6, 9.6             & 20.6, 15.9             & 22.5, 11.3 			    & 18.2, 9.1             & 16.1, 12.8              \\ \hline
		\textbf{$\gamma[\textup{ns}],\sigma_{U}[\textup{dB}]$}                             & 2.0, 5.2              & 9.2, 6.0           & 2.0, 5.0              & 9.9, 5.7 			    &    2.0, 4.6           &     2.4, 5.8           \\ \hline
		\textbf{$L_{\textup{AOD,max}},L_{\textup{AOA,max}}$}                                                        & 2, 2                     &  3, 3                   & 2, 2                     &  3, 3				    &       2, 2               &  2, 2                     \\ \hline
		\textbf{$\mu_{l,\textup{ZOD}}[\degree],\sigma_{l,\textup{ZOD}}[\degree]$}                 & -7.3, 3.8                 & -5.5, 2.9    & -7.2, 3.5                 & -5.8, 2.6                & -6.8, 4.9                   & -2.5, 2.7                    \\ \hline
		\textbf{$\mu_{l,\textup{ZOA}}[\degree],\sigma_{l,\textup{ZOA}}[\degree]$}                 & 7.4, 3.8                   & 5.5, 2.9     & 7.2, 3.5                   & 5.8, 2.6               & 7.4, 4.5                   & 4.8, 2.8                    \\ \hline
		\textbf{$\sigma_{\phi,\textup{AOD}}[\degree],\sigma_{\theta,\textup{AOD}}[\degree]$} & 7.1, 13.0          & 17.6, 13.0      & 7.0, 12.9          & 16.1, 12.9       & 4.3, 3.4             & 4.0, 3.3                \\ \hline
		\textbf{$\sigma_{\phi,\textup{AOA}}[\degree],\sigma_{\theta,\textup{AOA}}[\degree]$} & 19.3, 11.3          & 20.2, 11.6       & 19.9, 11.8          & 19.0, 11.6    & 4.4, 3.3             & 5.6, 3.3                \\ \hline
	\end{tabular}
\end{table*}

\section{Simulation Results} \label{sec:simulation}
The statistical channel model presented in Section \ref{sec:stat} was implemented in an indoor channel simulator based on NYUSIM outdoor channel simulator to investigate the accuracy of the simulated temporal and spatial statistics by comparing with the measured statistics. Note that the parameters listed in Table \ref{tab:values} are primary statistics used in the channel parameter generation procedure. The metrics used in this section for channel validation are secondary statistics such as RMS DS and RMS AS, which are not explicitly used in the channel generation, but the simulated and measured secondary statistics should yield good agreements. 10,000 simulations were carried out for each of four frequency scenarios (i.e., 28 GHz LOS, 28 GHz NLOS, 140 GHz LOS, and 140 GHz NLOS) presented in this work by generating 10,000 omnidirectional and directional PDPs, and 3-D AOD and AOA PASs as sample functions of (\ref{eq:omni_cir}) using the NYUSIM indoor channel simulator.
\subsection{Simulated RMS Delay Spreads}
\begin{figure}[h!]
	\centering
	\includegraphics[width=.8\linewidth]{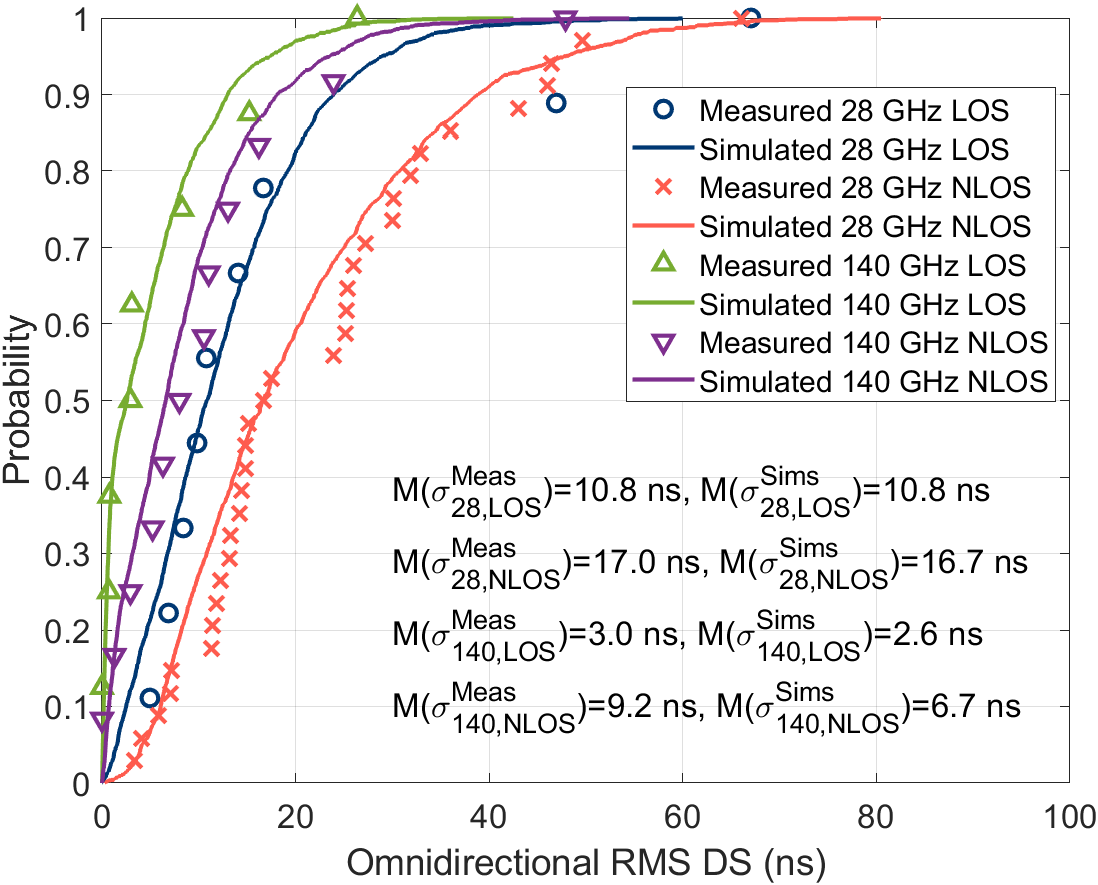}
	\caption{Omnidirectional RMS DS for 28 GHz and 140 GHz LOS and NLOS scenarios. Meas stands for measurement, and Sims stands for simulations.}
	\label{fig:rms_ds_4cases}
\end{figure}

The RMS DS describes channel temporal dispersion, a critical metric to validate a statistical channel model. Fig. \ref{fig:rms_ds_4cases} shows the simulated and measured omnidirectional RMS DS at 28 GHz and 140 GHz in LOS and NLOS scenarios. As shown in Fig. \ref{fig:rms_ds_4cases}, we obtained the empirical and simulated medians as 10.8 and 10.8 ns for the 28 GHz LOS scenario, 17.0 and 16.7 ns for the 28 GHz NLOS scenario, 3.0 and 2.6 ns for the 140 GHz LOS scenario, and 9.2 and 6.7 ns for the 140 GHz NLOS scenario. The simulated CDFs yield good agreements with the empirical CDFs for four frequency scenarios. 

\begin{figure}[h!]
	\centering
	\includegraphics[width=.8\linewidth]{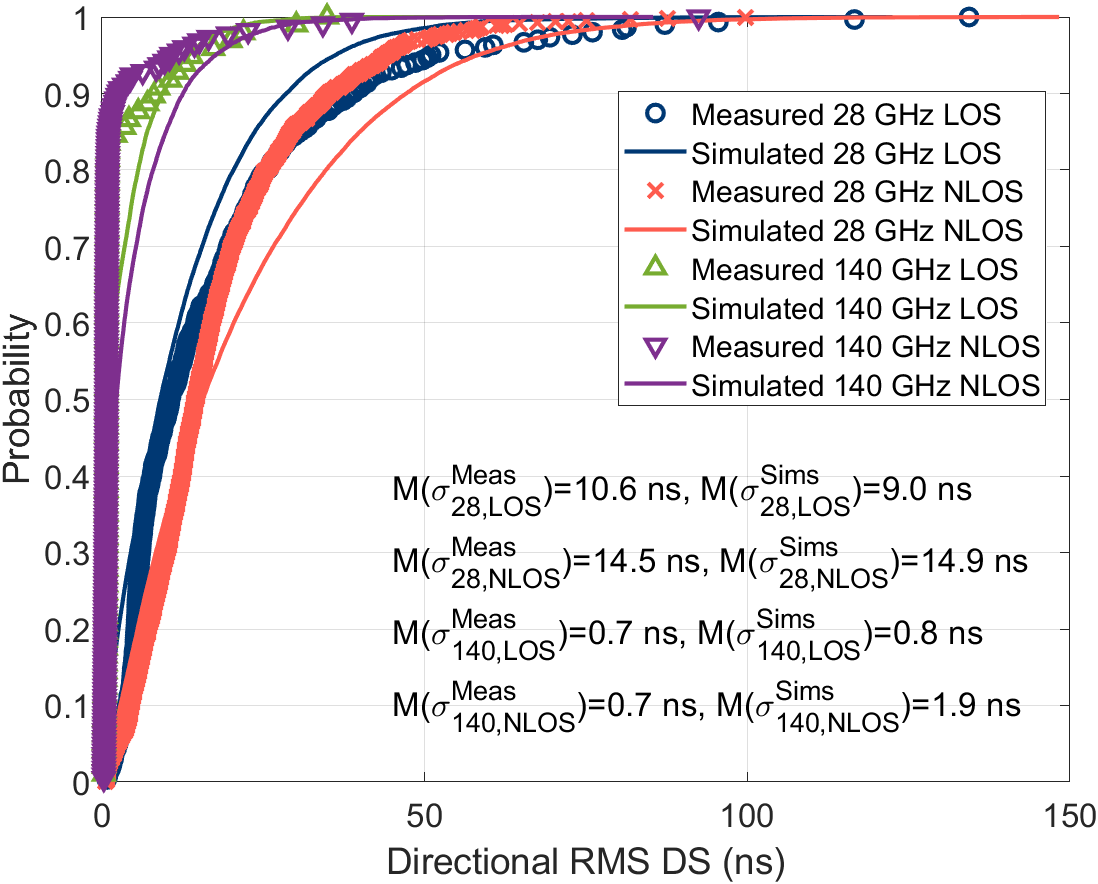}
	\caption{\textcolor{black}{Directional RMS DS for 28 GHz and 140 GHz LOS and NLOS scenarios. The simulated TX and RX antenna HPBWs for 28 GHz and 140 GHz in the azimuth and elevation plane are 30\degree~and 8\degree, respectively. }}
	\label{fig:dir_ds_4cases}
\end{figure}

By applying the antenna pattern, the directional CIR based on (\ref{eq:omni_cir}) can be given by
\begin{equation}
	\label{eq:dir_cir}
	\begin{split}
		h_{\text{dir}}(t,\Theta,\Phi) = &\sum_{n=1}^{N}\sum_{m=1}^{M_n}a_{m,n}e^{j\varphi_{m,n}}\cdot\delta(t-\tau_{m,n})\\
		&\cdot g_{\textup{TX}}(\overrightarrow{\Theta}-\overrightarrow{\Theta_{m,n}}) \cdot g_{\textup{RX}}(\overrightarrow{\Phi}-\overrightarrow{\Phi_{m,n}}),
	\end{split}
\end{equation} 
where $g_{\textup{TX}}(\overrightarrow{\Theta})$ and $g_{\textup{RX}}(\overrightarrow{\Theta})$ can be arbitrary 3-D TX and RX complex amplitude antenna patterns. The antenna pattern of horn antennas were used in directional CIR simulations in NYUSIM to compare with measured directional RMS DS from 28 GHz and 140 GHz measurements. The antenna gain of a horn antenna can be calculated using the given antenna HPBW, which was given by (45)-(46) in \cite{Samimi16mtt}. For each omnidirectional channel realization (i.e. PDP), the simulated horn antenna was pointing to the direction of each generated MPC, which output the same number of directional RMS DSs as the number of MPCs. The comparison between the measured and simulated directional RMS DS is shown in Fig. \ref{fig:dir_ds_4cases}. The simulated TX and RX antennas have 15 dBi gain with 30\degree~HPBW and 27 dBi gain with 8\degree~HPBW in both azimuth and elevation planes in 28 and 140 GHz channel simulations. \textcolor{black}{The measured directional RMS DSs in the LOS and NLOS scenarios are close with respect to both 28 GHz and 140 GHz. Furthermore, the median values of the measured and simulated directional RMS DS yield good agreements in the 28 GHz and 140 GHz LOS and NLOS scenarios.}

\subsection{Simulated RMS Angular Spreads}
\begin{figure}[h!]
	\centering
	\includegraphics[width=.8\linewidth]{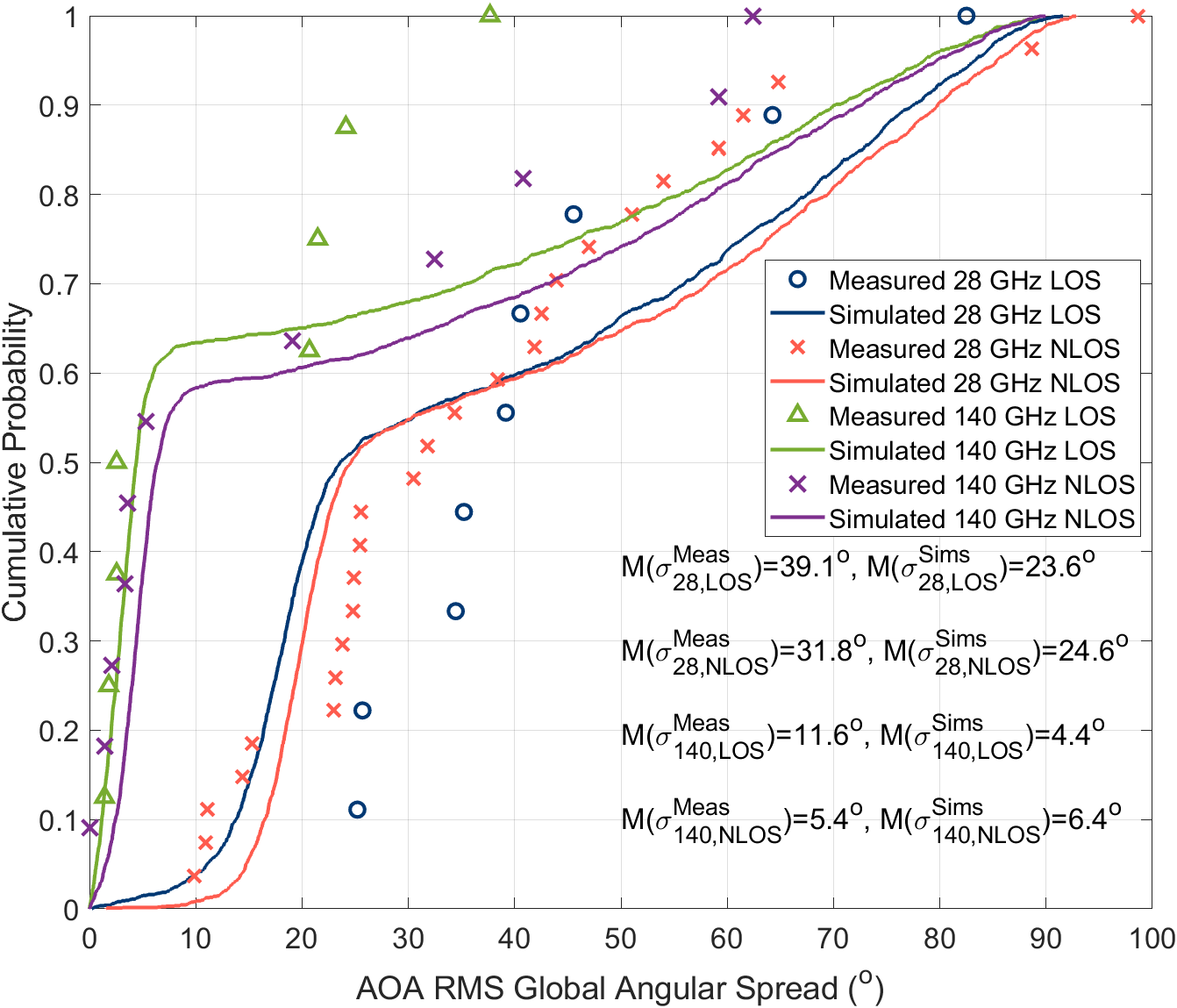}
	\caption{\textcolor{black}{Simulated and measured RMS global AOA AS for 28 and 140 GHz LOS and NLOS scenarios.} }
	\label{fig:rms_gas_4cases}
	\vspace{-4mm}
\end{figure}

\begin{figure}[h!]
	\centering
	\includegraphics[width=.8\linewidth]{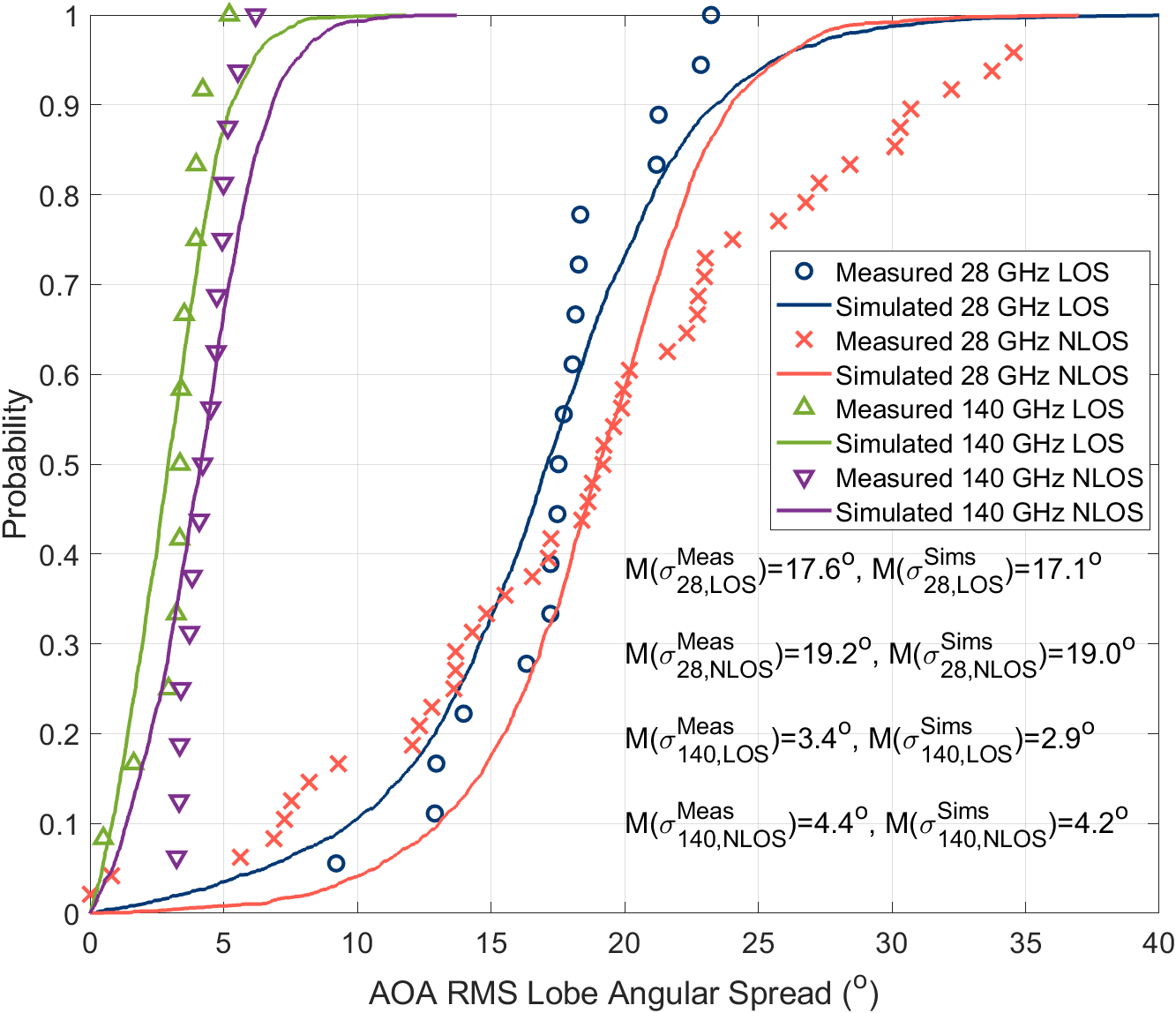}
	\caption{\textcolor{black}{Simulated and measured RMS lobe AOA AS for 28 and 140 GHz LOS and NLOS scenarios.}}
	\label{fig:rms_las_4cases}
\end{figure}

The omnidirectional azimuth and elevation AS describe the angular dispersion at a TX or RX over the entire 4$\pi$ steradian sphere, also termed global AS. The AOA and AOD global AS were computed using the total (integrated over delay) received power over all measured azimuth/elevation pointing angles. The measured and simulated global AOA RMS AS were calculated using Appendix A-1,2 in \cite{3GPP38901r16}. Fig. \ref{fig:rms_gas_4cases} shows that the simulated and measured median global ASs match well for 140 GHz but not well for 28 GHz due to the difference in the measured and simulated statistics of spatial lobes and the limited number of data samples. The simulated number of spatial lobes was uniformly distributed, which cannot perfectly recreate the specific statistics of spatial lobes measured in this environment, but may be well generalized to measurement data from various indoor office environments. The sheer increasing of the cumulative probability from 0 to 0.6 within 10\degree~at 140 GHz indicates only one SL (one main direction of arrival) observed at the RX, where the global RMS AS would be close to the lobe RMS AS as shown in Fig. \ref{fig:rms_las_4cases}.

The directional azimuth and elevation AS describe the degree of angular dispersion in a certain direction, which can be regarded as the lobe AS due to the definition of SLs. A -15 dB SLT was applied to obtain SLs before calculating the lobe RMS AS. The simulated and measured AOA RMS lobe AS for 28 and 140 GHz NLOS scenarios are compared in Fig. \ref{fig:rms_las_4cases}, where the median lobe AS of measured and simulated channels show an excellent agreement (within 0.5\degree). The measured lobe AS for 28 GHz is larger than the measured values for 140 GHz, which may be partly attributed to the difference in antenna HPBW (30\degree~and 8\degree~HPBW in 28 GHz and 140 GHz measurements, respectively). 

\section{Conclusion} \label{sec:conclusion}
The paper presented a 3-D spatial statistical channel model for mmWave and sub-THz frequencies in LOS and NLOS scenarios based on the extensive measurements at 28 and 140 GHz in an indoor office building. The omnidirectional and directional CI path loss models were derived from measurements, suggesting that NLOS propagation at both frequencies experience similar path loss over distance after removing the effect of the first meter of free space propagation loss. The extracted channel statistics showed that the number of TCs and the number of subpaths within each TC decrease as frequency increases. The channel generation procedure was listed step by step in Table \ref{tab:model}, and the values for required parameters obtained from 28 and 140 GHz LOS and NLOS measurements were given in Table \ref{tab:values}. The indoor channel simulator NYUSIM 3.0 based on the presented statistical model was used to generate tens of thousands of PDP and APS samples. The simulated secondary channel statistics (i.e., omnidirectional and directional RMS DS, global and lobe RMS AS) yielded good agreements with the measured channel statistics. The empirical channel statistics and corresponding unified statistical channel models across mmWave and sub-THz frequencies will provide insights for future propagation measurement and modeling in such frequency range and support analysis and design of 6G indoor wireless systems and beyond. 
\appendix
The processed data used to generate and calibrate the omnidirectional channel models in this paper are given in Table \ref{tab:28stat_new} and \ref{tab:140stat_new} as follows. 
\vspace{3mm}

\tablecaption{\textcolor{black}{28 GHz omnidirectional channel statistics with corresponding environment (Env.), TX IDs, RX IDs, T-R separation distance (T-R) in meters, path loss (PL) \cite{Mac15access} in dB, the number of TCs (\#TC), the number of SPs (\#SP), and the omnidirectional RMS DS in ns.}}

\tablefirsthead{\hline \textbf{Env.}&\textbf{\begin{tabular}[c]{@{}c@{}}TX \\ ID\end{tabular}}&\textbf{\begin{tabular}[c]{@{}c@{}}RX \\ ID\end{tabular}}&\textbf{\begin{tabular}[c]{@{}c@{}}T-R \\ (m)\end{tabular}}&\textbf{\begin{tabular}[c]{@{}c@{}}PL \\ (dB)\end{tabular}}&\textbf{\#TC}&\textbf{\#SP}&\textbf{\begin{tabular}[c]{@{}c@{}}DS \\ (ns)\end{tabular}}\\ \hline }
\label{tab:28stat_new}
\tablehead{%
	\hline
	\textbf{Env.}&\textbf{\begin{tabular}[c]{@{}c@{}}TX \\ ID\end{tabular}}&\textbf{\begin{tabular}[c]{@{}c@{}}RX \\ ID\end{tabular}}&\textbf{\begin{tabular}[c]{@{}c@{}}T-R \\ (m)\end{tabular}}&\textbf{\begin{tabular}[c]{@{}c@{}}PL \\ (dB)\end{tabular}}&\textbf{\#TC}&\textbf{\#SP}&\textbf{\begin{tabular}[c]{@{}c@{}} DS \\ (ns)\end{tabular}} \\ \hline}
\tabletail{%
	 \hline}
\begin{supertabular}{|c|c|c|c|c|c|c|c|}
	LOS& 1& 1& 6.4& 69.3& 4 & 11& 14.1 \\ \hline
	LOS& 1& 4& 7.9& 75.3& 4 & 9& 10.8 \\ \hline
	LOS& 1& 7& 12.9& 76.5& 5 & 11& 8.4 \\ \hline
	LOS& 2& 10& 4.1& 66.3& 4 & 8& 6.9 \\ \hline
	LOS& 3& 16& 5.3& 68.0& 3 & 8& 4.9 \\ \hline
	LOS& 4& 11& 12.7& 74.3& 9 & 31& 67.0 \\ \hline
	LOS& 4& 12& 7.1& 71.3& 5 & 20& 46.9 \\ \hline
	LOS& 4& 16& 20.6& 74.5& 4 & 18& 16.7 \\ \hline
	LOS& 4& 28& 21.3& 75.4& 3 & 18& 9.8 \\ \hline
	NLOS& 1& 2& 7.8& 76.6& 8 & 33& 14.4 \\ 
	NLOS& 1& 3& 10.1& 82.7& 3 & 19& 13.3 \\ \hline
	NLOS& 1& 5& 11.9& 84.3& 3 & 8& 4.1 \\ \hline
	NLOS& 1& 6& 14.4& 86.4& 5 & 21& 13.2 \\ \hline
	NLOS& 1& 8& 25.9& 95.9& 11 & 16& 46..0 \\ \hline
	NLOS& 2& 11& 9.0& 78.5& 6 & 15& 12.2 \\ \hline
	NLOS& 2& 12& 28.5& 89.5& 9 & 24& 49.7 \\ \hline
	NLOS& 2& 14& 30.4& 119.3& 5 & 48& 25.4 \\ \hline
	NLOS& 2& 15& 39.2& 115.8& 10 & 57& 66.0 \\ \hline
	NLOS& 2& 16& 41.9& 98.6& 12 & 32& 43.0 \\ \hline
	NLOS& 2& 18& 12.1& 93.6& 7 & 50& 26.0 \\ \hline
	NLOS& 2& 20& 17.7& 111.3& 8 & 43& 27.2 \\ \hline
	NLOS& 2& 21& 6.7& 78.0& 4 & 9& 11.3 \\ \hline
	NLOS& 3& 24& 7.8& 86.1& 9 & 43& 25.2 \\ \hline
	NLOS& 3& 25& 8.4& 76.3& 4 & 37& 15.2 \\ \hline
	NLOS& 3& 26& 5.5& 72.9& 3 & 5& 3.4 \\ \hline
	NLOS& 3& 27& 8.3& 75.8& 4 & 17& 7.0 \\ \hline
	NLOS& 4& 15& 20.8& 97.5& 4 & 40& 17.5 \\ \hline
	NLOS& 4& 18& 33.0& 96.8& 12 & 41& 46.5 \\ \hline
	NLOS& 5& 8& 3.9& 73.8& 5 & 10& 23.7 \\ \hline
	NLOS& 5& 19& 6.9& 75.3& 2 & 13& 5.8 \\ \hline
	NLOS& 5& 28& 15.6& 86.1& 4 & 18& 7.1 \\ \hline
	NLOS& 5& 29& 15.0& 81.3& 6 & 30& 31.9 \\ \hline
	NLOS& 5& 30& 11.4& 88.7& 9 & 28& 25.1\\ \hline
	NLOS& 5& 31& 13.9& 90.4& 2 & 14& 11.4 \\ \hline
	NLOS& 5& 32& 31.2& 90.2& 4 & 10& 16.6 \\ \hline
	NLOS& 5& 33& 9.1& 97.0& 7 & 32& 32.9 \\ 
\end{supertabular}%

\vspace{6mm}
\tablecaption{\textcolor{black}{140 GHz omnidirectional channel statistics with corresponding environment (Env.), TX IDs, RX IDs, T-R separation distance (T-R) in meters, path loss (PL) in dB, the number of TCs (\#TC), the number of SPs (\#SP), and the omnidirectional RMS DS in ns.}}
\tablefirsthead{\hline \textbf{Env.}&\textbf{\begin{tabular}[c]{@{}c@{}}TX \\ ID\end{tabular}}&\textbf{\begin{tabular}[c]{@{}c@{}}RX \\ ID\end{tabular}}&\textbf{\begin{tabular}[c]{@{}c@{}}T-R \\ (m)\end{tabular}}&\textbf{\begin{tabular}[c]{@{}c@{}}PL\\ (dB)\end{tabular}}&\textbf{\#TC}&\textbf{\#SP}&\textbf{\begin{tabular}[c]{@{}c@{}}DS \\ (ns)\end{tabular}}\\ \hline}
\label{tab:140stat_new}
\tablehead{%
	\hline
	\textbf{Env.}&\textbf{\begin{tabular}[c]{@{}c@{}}TX \\ ID\end{tabular}}&\textbf{\begin{tabular}[c]{@{}c@{}}RX \\ ID\end{tabular}}&\textbf{\begin{tabular}[c]{@{}c@{}}T-R \\ (m)\end{tabular}}&\textbf{\begin{tabular}[c]{@{}c@{}}PL\\ (dB)\end{tabular}}&\textbf{\#TC}&\textbf{\#SP}&\textbf{\begin{tabular}[c]{@{}c@{}}DS \\ (ns)\end{tabular}} \\ \hline}
\tabletail{%
	\hline }
\begin{supertabular}{|c|c|c|c|c|c|c|c|}
	LOS& 1& 1& 6.4& 88.8& 2& 3& 2.9  \\ \hline
	LOS& 1& 4& 7.9& 91.0& 2 & 3& 0.7 \\ 
	LOS& 1& 7& 12.9& 99.4& 2 & 4 & 3.1 \\ \hline
	LOS& 2& 10& 4.1& 88.9& 1 & 2 & 0.8 \\ \hline
	LOS& 3& 16& 5.3& 91.3& 1 & 1& 0 \\ \hline
	LOS& 4& 11& 12.7& 97.6& 2 & 5& 26.4 \\ \hline
	LOS& 4& 12& 7.1& 93.4& 2 & 2& 8.3 \\ \hline
	LOS& 4& 28& 21.3& 95.6& 3 & 9& 15.2 \\ \hline
	NLOS& 1& 2& 7.8& 103.2& 4 & 5& 2.9 \\ \hline
	NLOS& 1& 3& 10.1& 106.0& 2 & 5& 6.3 \\ \hline
	NLOS& 1& 5& 11.9& 105.1& 1 & 5& 1.3 \\ \hline
	NLOS& 1& 6& 14.4& 102.6& 3 & 4& 10.5 \\ \hline
	NLOS& 2& 11& 9.0& 108.0& 3 & 4& 13.0 \\ \hline
	NLOS& 2& 12& 28.5& 112.5& 3 & 10& 11.0 \\ \hline
	NLOS& 2& 15& 39.2& 114.4& 5 & 10& 23.9 \\ \hline
	NLOS& 2& 21& 6.7& 117.6& 3 & 3 & 5.2 \\ \hline
	NLOS& 5& 28& 6.4& 110.5& 2 & 3& 8.0 \\ \hline
	NLOS& 5& 31& 6.4& 139.8& 1 & 1& 0 \\ \hline
	NLOS& 5& 32& 6.4& 117.7& 4 & 6& 16.2 \\ \hline
	NLOS& 5& 33& 6.4& 113.2& 2 & 2& 47.9 \\ 
\end{supertabular}%

\bibliographystyle{IEEEtran}
\bibliography{jsac_final}

\end{document}